\documentclass[aps,prd,twocolumn,prd,showpacs,showkeys,preprintnumbers,bibnotes,floatfix,longbibliography,notitlepage,nofootinbib,superscriptaddress,superscriptgaddress,
]{revtex4-2}

\pdfoutput=1
\usepackage[colorlinks=true,breaklinks=true]{hyperref}
\usepackage{amsmath}
\usepackage{amsfonts}
\usepackage{amssymb}
\usepackage{mathrsfs}
\usepackage{graphicx}
\usepackage{color}
\usepackage[utf8]{inputenc}
\usepackage{amsfonts}
\usepackage{graphicx}
\usepackage{booktabs}
\usepackage{multirow}
\usepackage{makecell}
\usepackage{siunitx}
\usepackage{ragged2e}
\usepackage{rotating}
\usepackage{float}
\usepackage[dvipsnames]{xcolor}
\definecolor{darkred}{rgb}{0.5,0,0}
\definecolor{darkblue}{rgb}{0,0,0.5}
\definecolor{firebrick}{rgb}{0.75,0.125,0.125}
\definecolor{darkgreen}{rgb}{0,0.5,0}
\hypersetup{urlcolor=darkblue,
	    citecolor=darkgreen,
	    linkcolor=firebrick}

\usepackage{orcidlink}
\usepackage{lipsum}
\usepackage[normalem]{ulem}
\usepackage{enumitem}
\usepackage[capitalise]{cleveref}
\usepackage{pifont}
\usepackage{subfigure}
\usepackage{physics}
\usepackage{comment}

\long\def\exclude#1{}

\newcommand{\ie}{{\it i.e.}}

\newcommand{\eg}{{\it e.g.}}

\newcommand{\eq}{Eq.}

\newcommand{\fig}{Fig.}

\newcommand{\Refe}{Ref.}
\newcommand{\Refes}{Refs.}
\newcommand{\equ}[1]{\eq~(\ref{equ:#1})}
\newcommand{\figu}[1]{\fig~\ref{fig:#1}}

\graphicspath{{figures/}}

\begin{document}

\title{Identifying Energy-Dependent Flavor Transitions\\in High-Energy Astrophysical Neutrino Measurements}

\author{Qinrui Liu}
\email{qinrui.liu@queensu.ca}
\affiliation{Department of Physics, Engineering Physics and Astronomy, Queen's University, Kingston ON K7L 3N6, Canada}
\affiliation{Arthur B. McDonald Canadian Astroparticle Physics Research Institute,  Kingston ON K7L 3N6, Canada}
\affiliation{Perimeter Institute for Theoretical Physics, Waterloo ON N2L 2Y5, Canada}

\author{Damiano F.~G.~Fiorillo}
\email{damiano.fiorillo@nbi.ku.dk}
\affiliation{Niels Bohr International Academy, Niels Bohr Institute,\\University of Copenhagen, 2100 Copenhagen, Denmark}

\author{Carlos A. Arg{\"u}elles}
\email{carguelles@fas.harvard.edu}
\affiliation{Department of Physics \& Laboratory for Particle Physics and Cosmology, Harvard University, Cambridge, MA 02138, USA}

\author{Mauricio Bustamante}
\email{mbustamante@nbi.ku.dk}
\affiliation{Niels Bohr International Academy, Niels Bohr Institute,\\University of Copenhagen, 2100 Copenhagen, Denmark}

\author{Ningqiang Song}
\email{songnq@itp.ac.cn}
\affiliation{Institute of Theoretical Physics, Chinese Academy of Sciences, Beijing, 100190, China}
\affiliation{Department of Mathematical Sciences, University of Liverpool, Liverpool, L69 7ZL, United Kingdom}

\author{Aaron C. Vincent}
\email{aaron.vincent@queensu.ca}
\affiliation{Department of Physics, Engineering Physics and Astronomy, Queen's University, Kingston ON K7L 3N6, Canada}
\affiliation{Arthur B. McDonald Canadian Astroparticle Physics Research Institute,  Kingston ON K7L 3N6, Canada}
\affiliation{Perimeter Institute for Theoretical Physics, Waterloo ON N2L 2Y5, Canada}

\date{\today}

\begin{abstract}
 The flavor composition of TeV--PeV astrophysical neutrinos, \ie, the proportion of neutrinos of different flavors in their flux, is a versatile probe of high-energy astrophysics and fundamental physics.  Because flavor identification is challenging and the number of detected high-energy astrophysical neutrinos is limited, so far measurements of the flavor composition have represented an average over the range of observed neutrino energies.  Yet, this washes out the potential existence of changes in the flavor composition with energy and weakens our sensitivity to the many models that posit them.  For the first time, we measure the energy dependence of the flavor composition, looking for a transition from low to high energies.  Our present-day measurements, based on the 7.5-year public sample of IceCube High-Energy Starting Events (HESE), find no evidence of a flavor transition.  The observation of HESE and through-going muons jointly by next-generation neutrino telescopes Baikal-GVD, IceCube-Gen2, KM3NeT, P-ONE, TAMBO, and TRIDENT may identify a flavor transition around 200~TeV by 2030.  By 2040, we could infer the flavor composition with which neutrinos are produced with enough precision to establish the transition from neutrino production via the full pion decay chain at low energies to muon-damped pion decay at high energies.
\end{abstract}

\maketitle


\section{Introduction}
\label{sec:intro}

In astrophysics, neutrinos hold an esteemed position as compelling messengers from the cosmos.
Their minute interaction cross sections~\cite{Pauli:1930pc} make these elusive particles nearly impossible to stop or deflect en route to Earth, allowing them to convey astrophysical information directly from their sources, even if these are embedded in dense environments or produced cosmological distances away.
As a consequence, the study of neutrinos provides an invaluable and unique perspective on the most energetic phenomena in the Universe~\cite{Anchordoqui:2013dnh, Ahlers:2018fkn, Ackermann:2019ows, Meszaros:2019xej, Halzen:2019qkf, Palladino:2020jol, AlvesBatista:2021eeu, Ackermann:2022rqc, Guepin:2022qpl} and a handle on new physics beyond the reach of terrestrial experiments~\cite{Gaisser:1994yf, Ahlers:2018mkf, Ackermann:2019cxh, Arguelles:2019rbn, AlvesBatista:2021eeu, Ackermann:2022rqc, Arguelles:2022tki}. 

\begin{figure}[t!]
 \centering
 \includegraphics[width=\columnwidth]{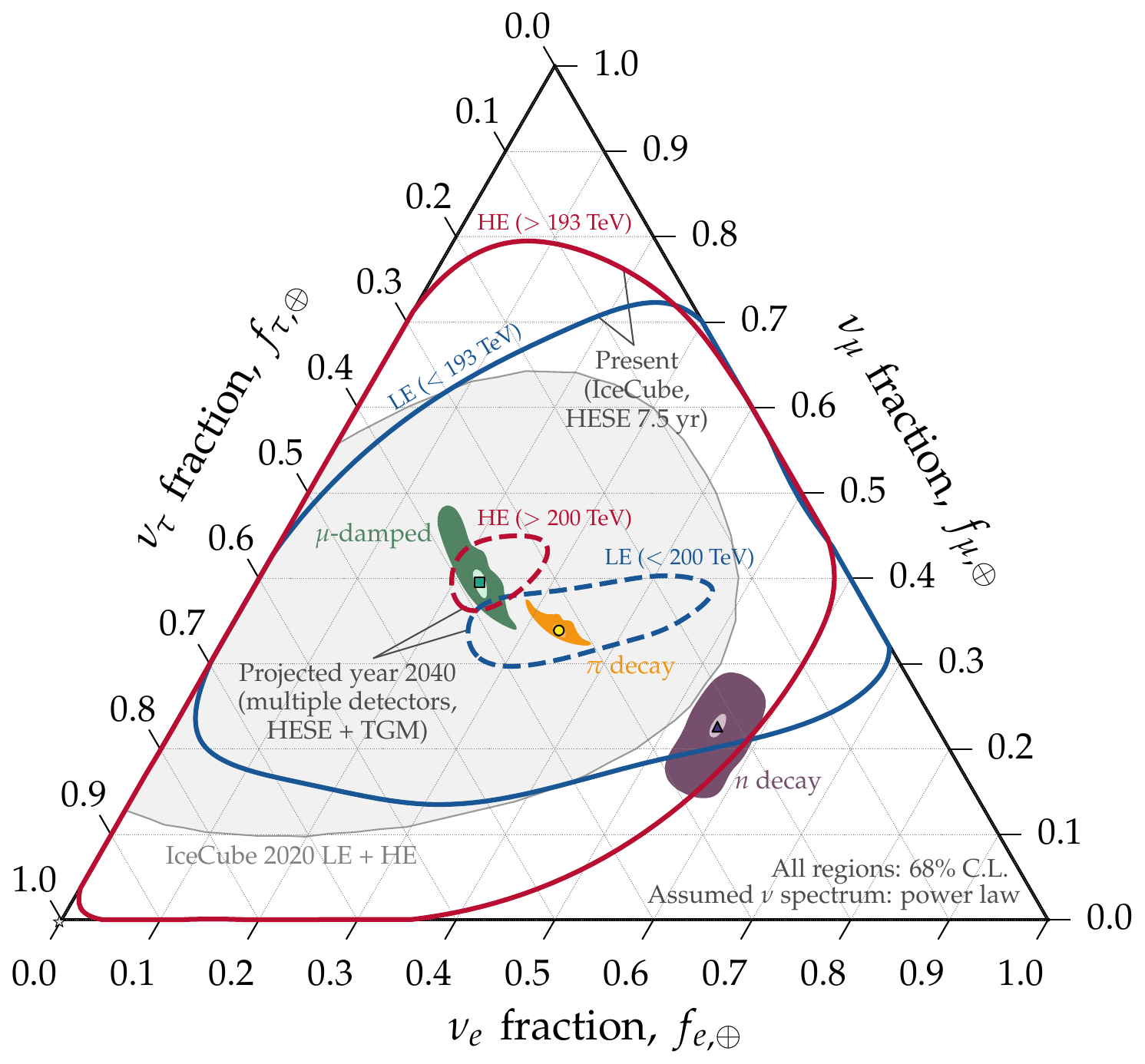}
 \caption{\textbf{\textit{Present and projected measurements of the high-energy neutrino flavor composition at low (LE) and high energy (HE).}}  Measurements are of the flavor composition at Earth, $f_{\alpha, \oplus}^{\rm LE}$ and $f_{\alpha, \oplus}^{\rm HE}$ ($\alpha = e, \mu, \tau$).  Present-day measurements are from the 7.5-year IceCube High-Energy Starting Event (HESE) sample~\cite{IceCube:2020wum, IC75yrHESEPublicDataRelease}.  Projections for the year 2040 are from HESE plus TGM detected by multiple planned detectors (Table~\ref{tab:detectors}).  In this figure, the neutrino spectrum is assumed to be a power law in energy. For comparison, we show the previous energy-averaged measurement by IceCube (``IceCube 2020 LE + HE)~\cite{IceCube:2020fpi}.  The allowed flavor regions for neutrino production via pion decay, muon-damped production, and neutron decay are computed as in \Refe~\cite{Song:2020nfh}, assuming present-day~\cite{Esteban:2020cvm, nufit5.1} and projected~\cite{Song:2020nfh} uncertainties in the neutrino mixing parameters.  The regions shown are at 68\% credibility level (C.L.).  See \figu{triangles} for other results and the main text for details.  \textit{Today, there is no evidence for a flavor transition in energy; by 2040, we could discover one.}
 }
 \label{fig:triangle_main}
\end{figure}

Over the past decade, IceCube~\cite{IceCube:2016zyt} and other neutrino observatories~\cite{BAIKAL:1997iok,AMANDA:2003dbe,ANTARES:2011hfw} have made significant strides in the field of high-energy neutrino astrophysics~\cite{Gaisser:1994yf,Learned:2000sw}. The groundbreaking discovery of a diffuse flux of high-energy neutrinos of astrophysical origin~\cite{IceCube:2013low,IceCube:2014stg,IceCube:2020wum} heralded the birth of high-energy neutrino astronomy. High-energy astrophysical neutrinos in the TeV-PeV range and beyond are believed to be produced in the most extreme environments, such as active galactic nuclei, starburst galaxies, and gamma-ray bursts. 
This has been confirmed by the recent observation of the first candidate astrophysical high-energy neutrino sources, the blazar TXS 0506+056~\cite{IceCube:2018dnn, IceCube:2018cha} and the Seyfert galaxy NGC 1068~\cite{IceCube:2022der}.
High-energy neutrinos can shed light on the physical processes taking place in these extreme environments and help to answer fundamental questions about the nature of the Universe~\cite{Arguelles:2019rbn}. Yet, our understanding of the sources and mechanisms of high-energy astrophysical neutrino production remains incomplete.

In this context, flavor-composition measurements have emerged as an essential tool in the arsenal of high-energy neutrino astrophysics.
Neutrinos exist in three flavors---electron ($\nu_e$), muon ($\nu_\mu$), and tau ($\nu_\tau$). They are produced with a certain flavor composition at the sources and undergo flavor oscillations during their journey from the sources to the detector. Therefore, the flavor of a single neutrino tells us very little about its history, but the flavor composition of the ensemble of detected neutrinos---\ie, the proportion of $\nu_e$, $\nu_\mu$, and $\nu_\tau$ in the diffuse neutrino flux that reaches Earth---carries crucial information about their production mechanisms, the identity of the neutrino sources, propagation effects, and even potential new physics~\cite{Learned:1994wg,Pakvasa:2007dc,Mena:2014sja,Palomares-ruiz:2015mka,Bustamante:2015waa,Arguelles:2015dca,Vincent:2016nut,Brdar:2016thq,Bustamante:2016ciw,Rasmussen:2017ert,Klop:2017dim,Farzan:2018pnk,Ahlers:2018yom,Bustamante:2018mzu,Arguelles:2019tum,Ahlers:2020miq,Arguelles:2021gwv,Carloni:2022cqz,Telalovic:2023tcb}. 

So far, flavor-composition measurements have faced two main challenges: a limited number of detected neutrinos and the difficulty in identifying the flavor of individual neutrino events~\cite{IceCube:2020fpi}. The interpretation of flavor measurements also suffers from uncertainties in the neutrino production and propagation. Thus, until now, such measurements have been done by averaging over a broad neutrino energy range~\cite{IceCube:2015gsk, Vincent:2016nut, IceCube:2020fpi}.
 
However, it is unlikely that in reality the flavor composition is completely energy-independent, since we expect different neutrino production mechanisms, with different yields of $\nu_e$, $\nu_\mu$, and $\nu_\tau$, to dominate at different energies. New physics effects could also modify flavor oscillations of neutrinos of different energies. 

In this work, we go beyond prior analyses by considering the prospects and challenges of measuring the high-energy astrophysical neutrino flavor composition if it is different at different neutrino energies. To do this, we consider three general benchmark scenarios for the shape of the neutrino energy spectrum, which is currently still uncertain: a single power law, a broken power law, and an abrupt change in spectrum normalization, each associated with a flavor transition from a low-energy (LE) value to a high-energy (HE) value.
We then examine the ability of neutrino telescopes to detect the existence of a flavor transition, measure the flavor composition at low and high energy, and from that infer the flavor composition at the sources~\cite{Bustamante:2019sdb, Song:2020nfh}. 

To do this, we first use existing IceCube data, and then produce forecasts based on next-generation neutrino telescopes, either proposed or under construction~\cite{Ackermann:2022rqc, Guepin:2022qpl}: Baikal-GVD~\cite{Baikal-GVD:2018isr}, IceCube-Gen2~\cite{IceCube-Gen2:2020qha}, KM3NeT~\cite{KM3Net:2016zxf}, P-ONE~\cite{P-ONE:2020ljt}, TAMBO~\cite{Thompson:2023pnl}, and TRIDENT~\cite{Ye:2022vbk}.
These telescopes will expand the cumulative rate of detection of high-energy neutrinos by over one order of magnitude.  We present forecasts based on information that has been publicly presented by the above experimental collaborations, including expected detector operation timelines, but point out that experimental configurations and timing will undoubtedly change during detector development, compressing or expanding these timelines. 

Figure~\ref{fig:triangle_main} illustrates our results for the measurement of the flavor composition at Earth; later, \figu{triangles} shows complete results.  We find that, today, there is no evidence of a flavor transition in energy in the sample of High-Energy Starting Events (HESE) detected by IceCube over 7.5 years~\cite{IceCube:2020wum, IC75yrHESEPublicDataRelease}.  However, our projections show sensitivity to detect a transition from pion decay at LE to muon-damped at HE in the coming years, using the combined detection of HESE and TGM events by the above upcoming neutrino telescopes.  These observations hold regardless of our choice of the shape of the neutrino spectrum from among our benchmarks.  Later, we show that this implies promising sensitivity to infer the flavor composition with which neutrinos are produced (\figu{posterior_src}). 

Our results pave the way toward a richer understanding of astrophysical high-energy neutrino production and enhanced searches for new high-energy neutrino physics.

The rest of this article is organized as follows. 
In \Cref{sec:scenarios} we describe the distinct astrophysical scenarios considered in this work.
In \Cref{sec:detectors} we describe the different experiments involved in our global analysis. 
In \Cref{sec:analysis} we describe our analysis techniques.  
In \cref{sec:results} we show our results for the measurement of the flavor composition at Earth and at the sources.
In \Cref{sec:improvements} we list envisioned improvements to our work.
In \Cref{sec:conclusion} we conclude.


\section{Astrophysical Scenarios}
\label{sec:scenarios}

\subsection{High-energy neutrino production}
\label{sec:scenarios_production}

Although the identities of the astrophysical sources responsible for the bulk of high-energy neutrinos detected are so far unknown, they are presumably cosmic accelerators---likely predominantly extragalactic---capable of boosting cosmic-ray protons and nuclei to energies of at least a few tens of PeV~\cite{Hillas:1984ijl, Anchordoqui:2018qom, AlvesBatista:2019tlv, Ackermann:2022rqc}. Upon interacting with matter~\cite{Margolis:1977wt, Stecker:1978ah, Kelner:2006tc} or radiation~\cite{Stecker:1978ah, Mucke:1999yb, Kelner:2008ke, Hummer:2010vx, Morejon:2019pfu}, these protons produce intermediate particles that decay to yield high-energy neutrinos.

High-energy neutrinos are believed to be primarily produced in the decay of pions via $\pi^-\to \mu^- + \bar{\nu}_\mu$, followed by $\mu^-\to e^-+\bar{\nu}_e+\nu_\mu$, and their charge-conjugated processes.  These pions are produced via photohadronic or hadronuclear processes of high-energy CRs and each neutrino receives about 5\% of the energy of the parent proton.
Therefore, from the full pion decay chain, the expected flavor composition of the neutrinos just after leaving the sources (S) is $(f_{e,{\rm S}}:f_{\mu,{\rm S}}:f_{\tau,{\rm S}})=(1:2:0)$, where $f_{\alpha, {\rm S}}$ ($\alpha = e, \mu, \tau$) is the ratio of $\nu_\alpha + \bar{\nu}_\alpha$ to the total flux.  Below, we denote by $\nu_\alpha$ the sum $\nu_\alpha + \bar{\nu}_\alpha$, unless otherwise indicated, since high-energy neutrino telescopes cannot presently distinguish between them.

The full pion decay chain yields the canonical expectation for the flavor composition of high-energy astrophysical neutrinos.  However, there are alternatives.  In sources that harbor intense magnetic fields, the intermediate muons might cool via synchrotron radiation before decaying.  In this muon-damped case, the only high-energy neutrinos are the $\nu_\mu$ produced directly in the decay of the pions, \ie, $(0:1:0)_{\rm S}$.  In addition, neutrinos may be produced in the beta-decay of neutrons, $n \to p + e^- + \bar{\nu}_e$, which yields a pure-$\bar{\nu}_e$ flux, \ie, $(1:0:0)_{\rm S}$, though these neutrinos may be more relevant at lower energies.

Other flavor compositions at the sources are possible (see, \eg, \Refes~\cite{Hummer:2010vx, Bustamante:2015waa}), but we adopt the three cases above---pion decay, muon-damped, and neutron decay---as benchmarks.  Later we explore transitions with neutrino energy between the first two (Section~\ref{sec:scenarios_transitions}).


\subsection{Flavor composition at Earth}
\label{sec:scenarios_oscillations}

After the high-energy neutrinos leave their sources, they travel cosmological-scale distances en route to Earth.  During that time, two effects modify the flavor content of their flux.  First, neutrinos oscillate in flavor; owing to their high energies, the oscillations vary rapidly with energy.  Second, the neutrino mass eigenstates, $\nu_1$, $\nu_2$, and $\nu_3$, whose superposition makes up the neutrino flavor states, separately.  As a result, high-energy neutrinos arrive at Earth as an incoherent sum of mass eigenstates, each containing a fraction $\lvert U_{\alpha i} \rvert^2$ of each flavor ($i = 1, 2, 3$), where $U$ is the Pontecorvo-Maki-Nakagawa-Sakata (PMNS) matrix, which depends on the mixing parameters $\theta_{12}$, $\theta_{23}$, $\theta_{13}$, and $\delta_{\rm CP}$, whose values are measured in oscillation experiments (Table~\ref{tab:parameters}).

Thus, for a given choice of the flavor composition at the sources, the flavor composition at Earth ($\oplus$) is~\cite{Learned:1994wg}
\begin{equation}
 \label{equ:flavor_ratios_earth}
 f_{\alpha,\oplus}
 =
 \sum_\beta f_{\beta, {\rm S}} 
 \sum_i 
 |U_{\alpha i}|^2 
 |U_{\beta i}|^2 \;.
\end{equation}
Even when the sum of mass eigenstates is not fully incoherent, the limited energy resolution of the neutrino detector makes it difficult to resolve the rapid energy dependence of the flavor oscillations, which leads again to \equ{flavor_ratios_earth} being the measurable flavor composition at Earth in practice.  New physics that modifies the flavor composition at Earth effectively modifies \equ{flavor_ratios_earth}.  Rather than exploring these deviations for specific new-physics models, later we explore classes of models generically by proxy, using benchmark neutrino spectra that include flavor transitions and that capture those deviations (\Cref{sec:scenarios_spectrum}).  In all cases, we adopt \equ{flavor_ratios_earth} in our analysis below to compute the flavor composition at Earth. 

We provide context for our measurements of the flavor composition by contrasting them against the allowed regions of flavor composition at Earth for our three benchmark choices of flavor composition at the sources (\Cref{sec:scenarios_production}) and, also, for any combination of flavor composition at the sources (Figs.~\ref{fig:triangle_main} and \ref{fig:triangles}).

We generate the allowed regions as in \Refe~\cite{Song:2020nfh} (see also \Refe~\cite{Bustamante:2015waa}), via \texttt{FANFIC}~\cite{FANFIC}.  To generate the present-day allowed flavor regions, we use the present-day values of the mixing parameters from the recent \texttt{NuFIT}~5.1~\cite{Esteban:2020cvm,nufit5.1} global fit to oscillation data (Table~\ref{tab:parameters}).  Evaluating the mixing parameters at their present-day best-fit values yields $\left(0.33:0.34:0.33\right)_\oplus$ for neutrino production via the full pion decay chain, $\left(0.23:0.39:0.38\right)_\oplus$ for muon-damped production, and $\left(0.55:0.23:0.22\right)_\oplus$ for production via neutron decay.  The present-day allowed regions of flavor composition at Earth---the \textit{theoretically palatable region}~\cite{Bustamante:2015waa}---are already relatively narrow, their sizes spanned mainly by the uncertainty in $\theta_{12}$ and $\theta_{23}$.  

By 2040, we expect the allowed flavor regions at Earth to shrink dramatically (Figs.~\ref{fig:triangle_main} and \ref{fig:triangles}), due to the foreseen improvement~\cite{Song:2020nfh} in precision in the measurement of the mixing parameters by upcoming experiments DUNE~\cite{Abi:2020wmh}, Hyper-Kamiokande~\cite{Abe:2018uyc}, and JUNO~\cite{An:2015jdp}.  Table~\ref{tab:parameters} shows that by 2040 $\mathrm{sin}^2\theta_{23}$ and $\mathrm{sin}^2\theta_{12}$ will be constrained to within $1\%$ precision, and that the correlation between $\mathrm{sin}^2\theta_{23}$ and $\delta_{\rm CP}$ that exists today~\cite{Esteban:2020cvm} will vanish.  

Later, in Section~\ref{sec:analysis_flavor-sources}, we use the present and projected distributions of the mixing parameters also to infer the flavor composition at the neutrino sources.


\subsection{Low-to-high-energy flavor transitions}
\label{sec:scenarios_transitions}

In general, the flavor composition might be different at different neutrino energies.  This could be due to effects from astrophysics---\ie, the dominant neutrino production mechanism changing with energy---or from neutrino physics---\ie, neutrino oscillations being affected by non-standard effects.  We survey these possibilities below.

Regarding astrophysics, a transition from pion-decay to muon-damped flavor composition from low to high neutrino energies could occur if the diffuse flux is due to a class of neutrino sources that harbor intense magnetic fields; see, \eg, \Refes~\cite{Kashti:2005qa, Lipari:2007su, Baerwald:2011ee, Bustamante:2020bxp, Fiorillo:2021hty, Bhattacharya:2023mmp}.  The magnetic field intensity in the sources determines the energy at which the flavor transition occurs, \ie, at which intermediate muons cool significantly via synchrotron radiation.  
A related possibility is that the diffuse flux originates from two different source populations, each emitting neutrinos with a different flavor composition and dominating in different energy ranges~\cite{Murase:2019vdl, Riabtsev:2022ynm}. 

Regarding neutrino physics, there is a large number of effects that could act during neutrino production, propagation, and detection to modify the flavor composition in an energy-dependent manner~\cite{Rasmussen:2017ert}.  During propagation alone, examples include neutrino decay~\cite{Beacom:2002vi, Mehta:2011qb, Baerwald:2012kc, Bustamante:2015waa, Shoemaker:2015qul, Bustamante:2016ciw, Denton:2018aml, Bustamante:2020niz, Abdullahi:2020rge}, Lorentz invariance violation~\cite{Barenboim:2003jm, Bustamante:2010nq, Arguelles:2015dca, Arguelles:2021kjg, Testagrossa:2023ukh, Telalovic:2023tcb}, dark matter interactions~\cite{Farzan:2018pnk, Karmakar:2020yzn, Arguelles:2023wvf}, non-standard interactions~\cite{Shoemaker:2015qul,Bustamante:2018mzu, Fiorillo:2020zzj, Berryman:2022hds, Agarwalla:2023sng}, sterile neutrinos~\cite{Brdar:2016thq, Arguelles:2019tum, Ahlers:2020miq, Arguelles:2021gwv}, pseudo-Dirac neutrinos~\cite{Beacom:2003eu, Shoemaker:2015qul, Carloni:2022cqz, Rink:2022nvw}, renormalization group-running of the neutrino mixing parameters~\cite{Bustamante:2010bf}, and the violation of the equivalence principle~\cite{Minakata:1996nd, Klop:2017dim, Fiorillo:2020gsb, Esmaili:2021omm, Chianese:2021vkf}.  

\begin{figure}[t!]
 \includegraphics[width=\columnwidth]{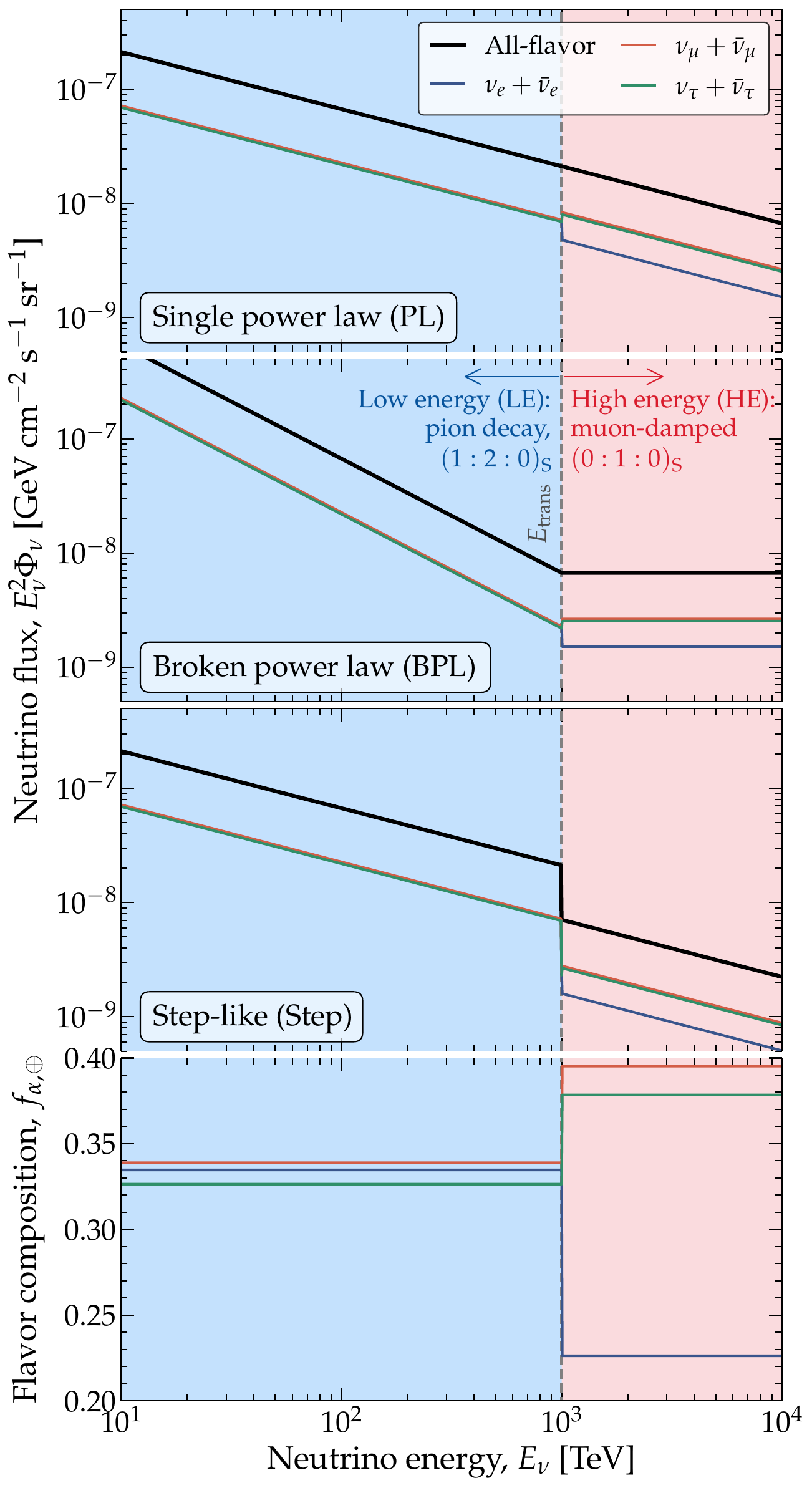}
 \caption{\textbf{\textit{Benchmark models of the high-energy neutrino spectrum at Earth.}}  The flavor composition, $f_{\alpha, \oplus}$, transitions at the energy $E_{\rm trans}$ from that expected from production via the full pion decay chain at LE to that expected from muon-damped production at HE.  For this plot, we compute the fluxes using the benchmark true values in Table~\ref{tab:parameters} and set $E_{\rm trans} = 1$~PeV; in our results, we consider also 200~TeV.  \textit{Top panels:} Neutrino spectra.  \textit{Bottom panel:} Flavor composition at Earth, common to the three benchmark spectra.  See Section~\ref{sec:scenarios} for details.  \textit{Our benchmark spectra capture the main features of a large variety of predicted changes to the flavor composition with energy.}}
 \label{fig:benchmark_flux}
\end{figure}

To assess the capability of identifying a flavor transition, we adopt a generic form of it that captures the essence of the transitions in the above models, rather than modeling each model separately.   We assume that the flavor composition of the observed diffuse neutrino flux transitions from a LE value, $\left(f_{e,\oplus}^{\rm LE}, f_{\mu,\oplus}^{\rm LE}, f_{\tau,\oplus}^{\rm LE} \right)$, to a HE value, $\left(f_{e,\oplus}^{\rm HE}, f_{\mu,\oplus}^{\rm HE}, f_{\tau,\oplus}^{\rm HE} \right)$, at an energy $E_{\rm trans}$.  

Later, in Sections~\ref{sec:analysis} and \ref{sec:results}, when analyzing present data, we make no assumptions on the values of these parameters, but rather extract them from the data.  When making projections, we assume that the flavor composition transitions from pion decay at LE to muon-damped at HE (see \Cref{sec:scenarios_oscillations} for the numerical values) and that $E_{\rm trans} = 200$~TeV or 1~PeV, and we assess how well we can recover those values from mock projected data.

Our choice of looking for a transition between the flavor composition from pion decay to that from muon-damped pion decay is conservative, since under standard oscillations they are fairly close to each other and, therefore, are challenging to distinguish; see Figs.~\ref{fig:triangle_main} and \ref{fig:triangles}.  In contrast, particular new-physics effects may introduce large deviations in the flavor composition relative to both of these benchmarks, making them easier to identify.

Often, changes to the flavor composition are accompanied by changes to the neutrino energy spectrum; next, we account for this.


\subsection{Neutrino energy spectrum}
\label{sec:scenarios_spectrum}

Figure~\ref{fig:benchmark_flux} shows the three benchmark choices of the neutrino spectrum that we adopt in our analysis, illustrated for $E_{\rm trans} =1$~PeV.  At that energy, the flavor composition transitions (see \Cref{sec:scenarios_transitions}).  For one of our benchmarks (PL below), this is the only change that occurs; for the remaining two (BPL and Step below), the flavor transition is accompanied by a transition in the shape of the all-flavor spectrum.
\begin{description}
 \item[Single power law spectrum (PL)]
 This is our barest prescription: a power law in neutrino energy, $\Phi_\nu \propto E_\nu^{-\gamma}$, with a spectral index $\gamma$ common to all flavors, and with an abrupt flavor transition at energy $E_{\rm trans}$, but no change in the energy spectrum.  This spectrum may mimic, \eg, the transition due to neutrino decay or Lorentz invariance violation, in which the energy spectrum is not necessarily affected by the onset of new-physics effects that alter the flavor composition.  
 \item[Broken power law spectrum (BPL)] 
  The spectrum is a power law that changes slope at the flavor transition, \ie, it is $\propto E_\nu^{-\gamma^{\rm LE}}$ below $E_{\rm trans}$ and $\propto E_\nu^{-\gamma^{\rm HE}}$ above it.   This spectrum may mimic, \eg, a scenario where the LE flux is dominated by sources that emit neutrinos with a steep energy spectrum, like NGC 1068~\cite{IceCube:2022der}, and the HE flux, by sources that emit neutrinos with a shallow energy spectrum, like TXS 0506+056~\cite{IceCube:2018dnn,IceCube:2018cha}.
 \item[Step-like spectrum (Step)] 
  The spectrum is a power law $\propto E^{-\gamma}$, whose all-flavor normalization has a step-like, abrupt change at the flavor transition.  This spectrum roughly approximates a more realistic transition between the pion-decay and muon-damped flavor compositions. Since, in the latter, only the neutrinos from the direct pion decay, which comprise one third of the total, have high energies, the flavor transition is accompanied by a factor-of-3 reduction in the flux. 
\end{description}

In our benchmark spectra above, the flavor transition is abrupt in energy.  In reality, the transition may be softer due to the distribution of the sources in redshift---which smears the energy spectrum with which neutrinos are emitted---or to the spread in the values of the parameters that affect neutrino production within the population of neutrino sources; \eg, \Refes~\cite{Hummer:2010vx, Baerwald:2011ee}.  However, limited energy resolution and event rates render detectors largely insensitive to the detailed dependence of the flavor transition with energy.  Hence, our choice of an abrupt transition, although possibly unrealistic, should not affect our results significantly.

Later, as part of our statistical analysis in Section~\ref{sec:analysis}, we vary the values of the parameters that dictate the spectrum shape and the flavor composition below and above the transition, generate mock samples of detected events, and contrast them to samples generated using the assumed true values of the parameters.  Table~\ref{tab:parameters} shows the free model parameters that we vary in our analysis; we defer further discussion to Section~\ref{sec:analysis}.


\section{Present and near-future neutrino telescopes}
\label{sec:detectors}

\subsection{High-energy neutrino telescopes}
\label{sec:detectors_detectors}

We focus on mainly optical ``plum pudding''-style neutrino telescopes, consisting of a large volume of natural water or ice, instrumented with strings of photomultiplier tubes (PMTs) designed to detect the Cherenkov light emitted primarily by the electromagnetic products of neutrino-nucleon deep-inelastic scattering (DIS)~\cite{IceCube:2017roe, Bustamante:2017xuy, Aartsen:2018vez, IceCube:2020rnc}.  
Depending on detector size and string spacing, these can be effective from neutrino energies of a few GeV to upward of 10--100 PeV.  (Above these energies, alternative detection strategies, such as in-ice Askaryan radio-detection, are more suitable~\cite{Ackermann:2022rqc, Barwick:2022vqt}.)

To produce our results, we use present-day data from IceCube, and projections for the planned optical-Cherenkov telescopes Baikal-GVD, IceCube-Gen2, KM3NeT, P-ONE, and TRIDENT.  Additionally, we consider TAMBO, a surface array of water-Cherenkov detectors that targets PeV-scale $\nu_\tau$.  We do not consider the recent proposal for a 30-km$^3$ in-water Cherenkov detector~\cite{LHASSOandNuTel, Huang:2023mzt}, since it is still preliminary.

A core assumption in making our predictions is that the capabilities of future detectors to detect HESE and TGM will be comparable to those of IceCube.  Neutrino detection at IceCube serves as the basis of our projections for detection in upcoming telescopes: as in \Refes~\cite{Song:2020nfh, Fiorillo:2022rft, Liu:2023lxz, Telalovic:2023tcb}, we estimate neutrino detection in each future telescope by scaling the rate of detected events in IceCube (Section~\ref{sec:detectors_event-types}) by the size of the detector relative to IceCube (Table~\ref{tab:detectors}).  Admittedly, this is a crude approximation that foregoes detailed modeling of the specific capabilities of upcoming detectors, made necessary by the present lack of publicly available simulations of the performance of upcoming telescopes to the same level of detail as it is available for IceCube.    

\begin{table}[t!]
 \begin{ruledtabular}  
  \caption{\label{tab:detectors}\textbf{\textit{High-energy neutrino telescopes considered in this work.}}  The detector sizes are given in units of the size of IceCube (IC), which is approximately 1~km$^3$.  This factor rescales the expected rates of IceCube HESE events, computed using the IceCube Monte Carlo sample, and of IceCube TGM events.  The start and end years are estimated.  For IceCube, the start year is for the final detector configuration (we do not account for detection during detector construction) and the end year is when IceCube-Gen2 is estimated to start operations.  See Section~\ref{sec:detectors_detectors} for details.} 
  \centering
  \renewcommand{\arraystretch}{1.3}
  \begin{tabular}{cccc}
   \multirow{2}{*}{Detector} & 
   \multirow{2}{*}{\makecell{Size relative\\to IceCube}} & 
   \multirow{2}{*}{\makecell{Start\\year}} &
   \multirow{2}{*}{\makecell{End\\year}} \\ 
   \\ \hline
   IceCube      & 
   1 IC         & 
   2011         & 
   2030         \\
   Baikal-GVD   &
   1.5 IC       & 
   2025         & 
   $\cdots$     \\
   KM3NeT       & 
   2.8 IC       & 
   2025         & 
   $\cdots$     \\
   IceCube-Gen2 & 
   8 IC         & 
   2030         & 
   $\cdots$     \\
   P-ONE        & 
   3.2 IC       & 
   2030         & 
   $\cdots$     \\
   TAMBO        & 
   0.5 IC       & 
   2030         & 
   $\cdots$     \\
   TRIDENT      & 
   7.5 IC       & 
   2030         & 
   $\cdots$     \\
  \end{tabular}
 \end{ruledtabular}  
\end{table}

Table~\ref{tab:detectors} summarizes our assumptions about the size and dates of operation of the detectors, both of which are based on current estimates and subject to change in the future.  For future detectors, we consider data-taking only after they reach their final sizes, which yields conservative results, since detection during their construction phases would add additional data.
\begin{description}
 \item[IceCube]
  The IceCube Neutrino Observatory, at the South Pole, has been operational for over ten years.  It consists of a cubic kilometer of clear ice with 86 vertical strings, each equipped with 60 optical modules that house PMTs to detect the Cherenkov light emitted by particle showers triggered by neutrino DIS~\cite{Aartsen:2016nxy}.  Section~\ref{sec:detectors_event-types} describes the types of events detected by IceCube that we use in our analysis---HESE and TGM---and how we model neutrino detection inasmuch detail as possible outside the Collaboration.  
 \item[IceCube-Gen2]
  IceCube-Gen2 is the planned extension of IceCube~\cite{IceCube:2019pna,IceCube-Gen2:2020qha}.
  It will expand the detector by adding 120 new PMT strings, resulting in an instrumented volume of 7.9~km$^3$, yielding an effective area that is 7 to 8.5 times larger than IceCube between 100~TeV and 1~PeV.
  The deployment of IceCube-Gen2 is scheduled to start in 2027 and aims for completion by 2033.
 \item[KM3NeT]
  KM3Net~\cite{KM3Net:2016zxf}, the successor to ANTARES~\cite{ANTARES:2011hfw}, is an in-water Cherenkov detector under construction in the Mediterranean Sea.
  It is split into two sites, one densely-instrumented low-energy site, ORCA, and a larger and sparser site optimized for higher energies, ARCA; we consider only the latter in our analysis (heretofore, we refer to it simply as KM3NeT).
  ARCA will consist of two clusters of 115 strings, each equipped with 18 optical modules per string, situated 100~km off the coast of Sicily.
  The construction is underway and is expected to be finalized by 2027. 
  With an estimated event rate of 15.6 cascades (Section~\ref{sec:detectors_event-types}) induced by cosmic neutrinos per year, KM3Net is projected to have an event rate approximately 2.8 times that of IceCube.  We adopt this as a benchmark, but the exact figure depends on the shape of the astrophysical neutrino energy spectrum. 
 \item[Baikal-GVD]
  Baikal-GVD~\cite{Safronov:2020dtw}, in Lake Baikal, Siberia, the successor to the Baikal detector~\cite{BAIKAL:1997iok}, is an in-water Cherenkov detector, in operation since 2018.  Its current effective volume is 0.35~km$^3$; by 2025, when its construction is completed, it will reach 1.5~km$^3$.
  Baikal-GVD consists of 90 strings, each equipped with 12 optical modules.
  It has already observed~\cite{Baikal-GVD:2022fmn}, tentatively, neutrinos from the blazar TXS 0506+056 previously observed by IceCube~\cite{IceCube:2018cha, IceCube:2018dnn} and the diffuse flux of high-energy astrophysical neutrinos~\cite{Baikal-GVD:2022fis}.
 \item[P-ONE]
  P-ONE, the Pacific Ocean Neutrino Experiment~\cite{Agostini:2020aar}, is a planned in-water Cherenkov experiment to be located off Vancouver Island in the Cascadia basin.
  By 2023, a first string has been deployed and six strings are funded and are under construction. 
  The final configuration of P-ONE is expected to be completed by 2030 and will comprise 70 PMT strings, each equipped with 20 optical modules, covering a cylindrical volume with a height of 1~km and a radius of 1~km.
 \item[TRIDENT]
  TRIDENT~\cite{Ye:2022vbk}, the Tropical Deep-Sea Neutrino Telescope, is a proposed in-water Cherenkov telescope to be located in the South China Sea.
  The planned final configuration of TRIDENT consists of 1211 PMT strings yielding an effective volume of 7.5~km$^3$, TRIDENT aims to begin construction in 2026 and reach its full size by 2030.
 \item[TAMBO]
  TAMBO~\cite{Romero-Wolf:2020pzh, Thompson:2023pnl}, the Tau Air-Shower Mountain-Based Observatory, is a proposed surface array of water-Cherenkov detectors designed to detect PeV-scale $\nu_\tau$ via the upward-going tau showers they trigger, with an expected event rate of 7 $\nu_\tau$ per year. 
  Even though TAMBO is small compared to IceCube (Table~\ref{tab:detectors}), its particular sensitivity to $\nu_\tau$ provides unique, complementary information to plum-pudding-style optical neutrino telescopes, which are more evenly sensitive to all neutrino flavors.  We assume a start date of TAMBO of 2030. We include the contribution of TAMBO by considering only one third of its proposed volume from Table~\ref{tab:detectors}, to account for it being sensitive only to $\nu_\tau$. Since the neutrino detection strategy of TAMBO is different from the other neutrino telescopes discussed in this work, modeling it as a scaled-down version of IceCube is a particularly coarse approximation, but given its relative size the effect of mismodeling it on our results is small.
\end{description}


\subsection{Types of detected events}
\label{sec:detectors_event-types}

High-energy neutrino telescopes detect neutrinos primarily via neutrino deep-inelastic scattering (DIS) on nucleons, $N$.  A DIS interaction can be either charged-current (CC), \ie, $\nu_\alpha + N \to l_\alpha + X$, with $X$ final-state hadrons, or neutral-current (NC), \ie, $\nu_\alpha + N \to \nu_\alpha + X$.  At these energies, the DIS cross sections of $\nu_\alpha$ and $\bar{\nu}_\alpha$ are nearly equal, and we do not distinguish between events made by one or the other. The final-state hadrons receive, on average, 20\% of the energy the initial-state neutrino; the final-state lepton receives the rest.  

Charged final-state particles initiate high-energy showers; as they develop, they emit Cherenkov light that propagates through the ice and is detected by the PMTs.  From the spatial and temporal distribution of the detected light, \ie, from the \textit{event morphology}, analyses infer the energy and direction of the interacting neutrino.  From the relative number of events with different morphologies, analyses infer the flavor composition of the incoming flux of neutrinos.

To measure the flavor composition, we use two classes of detected events in high-energy neutrino telescopes: HESE---made by neutrinos of all flavors---and TGM --made mainly by $\nu_\mu$. 


\subsubsection{High-Energy Starting Events (HESE)}
\label{sec:detectors_event-types-hese}

HESE events are triggered by the interaction of neutrinos inside the instrumented detector volume.  Different neutrino flavors interacting via CC and NC generate three distinct HESE event morphologies: cascades, tracks, and double cascades.  The flavor composition of the neutrino flux responsible for a sample of detected HESE events is inferred by comparing the numbers of detected events of each morphology.
  
\begin{description}
 \item[Cascades]
  Cascades are generated primarily by the CC interaction of $\nu_e$. In one such interaction, the showers initiated by the final-state electrons and hadrons are superimposed and treated as one.  Cascades are also generated by the CC interaction of $\nu_\tau$ that produce short-lived final-state tau leptons whose decay length is shorter than the spacing between detector modules, such that the shower it initiates is indistinguishable from that initiated by the hadrons.  Finally, cascades are also generated by NC interactions of $\nu_e$, $\nu_\mu$, and $\nu_\tau$.  At a given energy, the NC contribution is subdominant because the NC cross section is smaller than the CC cross section and NC interactions deposit less energy into the shower, since only the final-state hadrons shower, though this is partially compensated by the fact that all flavors contribute.
 \item[Tracks]
  Tracks are generated by the CC interaction of $\nu_\mu$ and by the 17\% of CC interactions of $\nu_\tau$ that produce a final-state muon~\cite{ParticleDataGroup:2022pth}. In both cases, energetic muons propagate over tens of kilometers while losing energy, leaving tracks of Cherenkov light that are easily identifiable.
 \item[Double cascades]
  Double cascades are generated by the CC interaction of energetic $\nu_\tau$~\cite{Learned:1994wg}.  A first cascade initiated by the final-state hadrons is followed by a second cascade initiated by the hadronic decay of the final-state tau lepton.  Double cascades are identifiable for $\nu_\tau$ that are sufficiently energetic to yield final-state tau leptons whose decay length exceeds the spacing between detector modules.  Recently, IceCube observed the first two candidate double cascades~\cite{IceCube:2020fpi}.  The requirement of needing to observe the two cascades makes double cascades rare in the sample of detected events, but because they are clean evidence of $\nu_\tau$ they improve the measurement of the flavor composition appreciably~\cite{IceCube:2020fpi}.
\end{description} 

The contribution of neutrino-electron interactions to the event rate is ordinarily negligible, except within a narrow energy range around the Glashow resonance~\cite{Glashow:1960zz} at $E_\nu = 6.3$~PeV.  There, the scattering of $\bar{\nu}_e$ on electrons is enhanced significantly by the production of an on-shell $W$ boson.  Recently, IceCube detected the first candidate Glashow resonance event~\cite{IceCube:2021rpz}.  Detecting the Glashow resonance could allow us to break the degeneracy between events due to $\nu_e$ and $\bar{\nu}_e$~\cite{Biehl:2016psj, Huang:2019hgs, Bustamante:2020niz, Huang:2023yqz, Liu:2023lxz}.

The description of the event morphologies above exposes the core challenge in measuring the flavor composition: cascades and tracks can be made by more than one neutrino flavor, so it is not possible to firmly infer the flavor of any individual detected event.  This is aggravated by the limited energy and angular resolution of the detector, and by the fact that morphologies are occasionally mis-identified~\cite{Palomares-ruiz:2015mka, IceCube:2015rro}.

We account for these nuances by basing our computation of HESE event rates on the publicly available IceCube Monte Carlo (MC) sample of HESE events~\cite{IC75yrHESEPublicDataRelease}, released together with 7.5~year HESE sample~\cite{IceCube:2020wum}.  With it we compute separately the rates of HESE cascades---including from the Glashow resonance---tracks, and double cascades.  The MC sample accounts for the attenuation of the neutrino flux due to neutrino interactions with matter as it propagates inside the Earth, and implicitly contains a detailed description of IceCube, including of the relation between different flavors and morphologies.  We defer to \Refes~\cite{IceCube:2020wum, IC75yrHESEPublicDataRelease} and present a short overview in Section~\ref{sec:analysis}.


\subsubsection{Through-going muons (TGM)}
\label{sec:detectors_event-types-tgm}

A muon of energy larger than TeV can travel hundreds to thousands of meters, eventually crossing the neutrino detector and leaving an identifiable track of Cherenkov light in its wake.  The interacting muon can be of atmospheric origin, created in cosmic-ray-induced air showers, or it can be created in a $\nu_\mu$ CC interaction in the ice.  In the latter case, the long travel range of high-energy muons means that tracks can be detected even if the neutrino interactions that birth them happen far away, thus effectively extending the size of the detector.

In our analysis, we consider TGM, an event sample made up of tracks induced by CC interactions of $\nu_\mu$ outside the detector.  In TGM samples, cascade-like events are filtered out.  To circumvent the overwhelming background of atmospheric muons, our TGM samples are limited to up-going events, \ie, events that reach the detector from below it.  As a result, in our samples, up-going atmospheric muons are stopped inside Earth before reaching the detector, leading to a neutrino purity larger than 99.8\%~\cite{IceCube:2021uhz}. 

Including TGM events in our analysis vastly increases the number of events.  For comparison, IceCube detected 650,000 TGM events in 9.5 years~\cite{IceCube:2021uhz} \textit{vs.}~102 HESE events in 7.5 years~\cite{IceCube:2020wum}, with the caveat that most TGM events out of those are of atmospheric origin.  We combine TGM only in our projections, as explained below.
For the measurement of the flavor composition, including TGM events improves the measurement precision of the $\nu_\mu$ fraction; we show this explicitly in \Cref{sec:results_flavor-earth}.  

Unlike our HESE track samples, in our TGM samples we only include the contribution from muons made by $\nu_\mu$, not by $\nu_\tau$, since this is a recent development in the point-source searches that we base our samples off of.


\subsubsection{Combining HESE and TGM events}
\label{sec:detectors_event-types_hese-tgm}

By themselves, HESE events provide sensitivity to the flavor composition.  In our present-day results, we use HESE events exclusively.  However, their paucity limits the precision of the flavor-composition measurement.

To address this, in our projections we combine HESE events with TGM.  Our strategy is similar to \Refes~\cite{IceCube:2015gsk, Naab:2023xcz}, which, however, measured the flavor composition assuming that it is constant in energy.  Because vastly more TGM are detected than HESE events, combining them tightens the measurement of the muon-flavor content significantly.  

\section{Methods}
\label{sec:analysis}

\begingroup
\squeezetable
\begin{table*}[t!]
 \begin{ruledtabular}  
  \caption{\label{tab:parameters}\textbf{\textit{Free model parameters considered in this work.}}  The spectrum shape parameters define the neutrino energy spectrum in our three benchmark flux models (\figu{benchmark_flux}): single power law (PL), broken power law (BPL), and step-like (Step).  The transition from a pion-decay flavor composition at LE to a muon-damped flavor composition at HE takes place at neutrino energy $E_{\rm trans}$.  In our statistical analysis, the parameter values are varied as within the ranges given in the table.  See Section~\ref{sec:scenarios} for details on the flux models.  See Section~\ref{sec:analysis_flavor-earth} for details on the parametrization of the flavor composition at Earth used when inferring the flavor composition at the sources.} 
  \centering
  \renewcommand{\arraystretch}{1.35}
  \begin{tabular}{cccccccc}
   \multirow{2}{*}{Parameter}                          & 
   \multirow{2}{*}{Symbol}                             &
   \multirow{2}{*}{Units}                              &
   \multicolumn{3}{c}{Used in flux model}              &
   \multirow{2}{*}{\makecell{True value\\(in proj.)\footnote{\label{fnote1}The true parameters values are used only when making projections, to generate mock samples of observed events.}}} & 
   \multirow{2}{*}{Prior}                              \\
   &
   &
   &
   PL   &
   BPL  &
   Step &
   &
   \\
   \hline
   \multicolumn{8}{c}{Spectrum shape parameters (Section~\ref{sec:scenarios_spectrum})} \\
   \hline
   \multirow{2}{*}{\makecell{All-flavor flux normalization at 100~TeV,\\common to LE and HE}}                       &
   \multirow{2}{*}{$\Phi_{\nu,0}$}               &
   \multirow{2}{*}{$10^{-18}$~GeV$^{-1}$~cm$^{-2}$~s$^{-1}$~sr$^{-1}$}     &
   \multirow{2}{*}{$\checkmark$}                             &
   \multirow{2}{*}{$\checkmark$}                             &
   \multirow{2}{*}{}                                     &
   \multirow{2}{*}{$6.7$}                          &
   \multirow{2}{*}{Uniform $\in [0,10]$}           \\
   \\
   LE all-flavor flux normalization at 100~TeV     &
   $\Phi_{\nu,0}^{\rm LE}$               &
   $10^{-18}$~GeV$^{-1}$~cm$^{-2}$~s$^{-1}$~sr$^{-1}$               
   &
   &
   &
   $\checkmark$                                 &
   $6.7$                                    &
   Uniform $\in [0,10]$                     \\
   HE all-flavor flux normalization at 100~TeV     &
   $\Phi_{\nu,0}^{\rm HE}$                 &
   $10^{-18}$~GeV$^{-1}$~cm$^{-2}$~s$^{-1}$~sr$^{-1}$                &
   &
   &
   $\checkmark$                                 &
   $(6.7/3)$                                &
   Uniform $\in [0,10]$                     \\
   \multirow{2}{*}{Energy of flavor transition, LE to HE}    &
   \multirow{2}{*}{$E_{\rm trans}$}         &
   \multirow{2}{*}{TeV}                     &
   \multirow{2}{*}{$\checkmark$}                &
   \multirow{2}{*}{$\checkmark$}                &
   \multirow{2}{*}{$\checkmark$}                &
   \multirow{2}{*}{\makecell{200\\or $10^3$}} &
   \multirow{2}{*}{\makecell{Log$_{10}$-uniform \\ 
    $\in [60,10^4]$} }    \\
   \\
   Spectral index, common to LE and HE      &
   $\gamma$                                 &
   $\cdots$                                 &
   $\checkmark$                                 &
   &
   $\checkmark$                                 &
   2.5                                      &
   Uniform $\in [1,4]$                      \\
   LE spectral index                        &
   $\gamma^{\rm LE}$                        &
   $\cdots$                                 &
   &
   $\checkmark$                                 &
   &
   3.0                                      &
   Uniform $\in [1,4]$                      \\
   HE spectral index                        &
   $\gamma^{\rm HE}$                        &
   $\cdots$                                 &
   &
   $\checkmark$                                 &
   &
   2.0                                      &
   Uniform $\in [1,4]$                      \\
   \hline
   \multicolumn{8}{c}{Additional parameters used when measuring the flavor composition at Earth (Section~\ref{sec:analysis_flavor-earth})} \\   
   \hline
   LE angle of flavor composition at Earth  & 
   $\sin^4 \theta_\oplus^{\rm LE}$          &
   $\cdots$                                 &
   $\checkmark$                                 &
   $\checkmark$                                 &
   $\checkmark$                                 &
   0.45                                     &
   Uniform $\in [0,1]$                      \\
   LE angle of flavor composition at Earth  & 
   $\cos 2 \psi_\oplus^{\rm LE}$            &
   $\cdots$                                 &
   $\checkmark$                                 &
   $\checkmark$                                 &
   $\checkmark$                                 &
   -0.01                                    &
   Uniform $\in [-1,1]$                     \\
   HE angle of flavor composition at Earth  & 
   $\sin^4 \theta_\oplus^{\rm HE}$          &
   $\cdots$                                 &
   $\checkmark$                                 &
   $\checkmark$                                 &
   $\checkmark$                                 &
   0.39                                     &
   Uniform $\in [0,1]$                      \\
   HE angle of flavor composition at Earth  & 
   $\cos 2 \psi_\oplus^{\rm HE}$            &
   $\cdots$                                 &
   $\checkmark$                                 &
   $\checkmark$                                 &
   $\checkmark$                                 &
   -0.27                                    &
   Uniform $\in [-1,1]$                     \\
   \hline
   \multicolumn{8}{c}{Additional parameters used when inferring the flavor composition at the sources (Section~\ref{sec:analysis_flavor-sources})}   \\
   \hline
   LE electron flavor fraction              &
   $f_{e, \oplus}^{\rm LE}$                 &
   $\cdots$                                 &
   $\checkmark$                                 &
   $\checkmark$                                 &
   $\checkmark$                                 &
   0.33                                     &
   Uniform $\in [0,1]$                      \\
   HE electron flavor fraction              &
   $f_{e, \oplus}^{\rm HE}$                 &
   $\cdots$                                 &
   $\checkmark$                                 &
   $\checkmark$                                 &
   $\checkmark$                                 &
   0.23                                     &
   Uniform $\in [0,1]$                      \\
   \multirow{2}{*}{Solar mixing angle}      &
   \multirow{2}{*}{$\sin^2 \theta_{12}$}    &
   \multirow{2}{*}{$\cdots$}                &
   \multirow{2}{*}{$\checkmark$}                &
   \multirow{2}{*}{$\checkmark$}                &
   \multirow{2}{*}{$\checkmark$}                &
   \multirow{2}{*}{0.304}                   &
   Present\footnote{\label{fnote2}We build the present-day likelihood of the mixing parameters from the $\Delta \chi^2$ distributions of the \texttt{NuFit}~5.1 global fit to oscillation data.  We treat $\sin^2 \theta_{12}$ and $\sin^2 \theta_{13}$ as uncorrelated with the others, and $\sin^2 \theta_{23}$ and $\delta_{\rm CP}$ as correlated between them via a two-dimensional joint likelihood.  In the table, for brevity, we show only the approximate one-dimensional 68\%~credible intervals, but our results are generated using the detailed shape of the \texttt{NuFIT} distributions.}: $0.304 \pm 0.012$  \\
   &
   &
   &
   &
   &
   &
   &
   Proj.\footnote{\label{fnote3}All of the year-2040 projected likelihoods of the mixing parameters are one-dimensional normal distributions centered at their present-day best-fit values from \texttt{NuFIT}~5.1, with standard deviation $\sigma$.  There is no correlation between parameters; see \Refe~\cite{Song:2020nfh}.  The projected uncertainties shown in the table apply to the normal and inverted neutrino mass ordering.}: Normal, $\sigma = 0.002$ \\
   \multirow{2}{*}{Atmospheric}             &
   \multirow{2}{*}{$\sin^2 \theta_{23}$}    &
   \multirow{2}{*}{$\cdots$}                &
   \multirow{2}{*}{$\checkmark$}                &
   \multirow{2}{*}{$\checkmark$}                &
   \multirow{2}{*}{$\checkmark$}                &
   \multirow{2}{*}{0.450}                   &
   Present\footref{fnote2}: $0.450_{-0.019}^{+0.016}$  \\
   &
   &
   &
   &
   &
   &
   &
   Proj.\footref{fnote3}: Normal, $\sigma = 0.004$ \\   
   \multirow{2}{*}{Reactor mixing angle}      &
   \multirow{2}{*}{$\sin^2 \theta_{13}$}    &
   \multirow{2}{*}{$\cdots$}                &
   \multirow{2}{*}{$\checkmark$}                &
   \multirow{2}{*}{$\checkmark$}                &
   \multirow{2}{*}{$\checkmark$}                &
   \multirow{2}{*}{0.304}                   &
   Present\footref{fnote2}: $0.02246 \pm 0.00062$  \\
   &
   &
   &
   &
   &
   &
   &
   Proj.\footref{fnote3}: Normal, $\sigma = 0.00062$ \\   
   \multirow{2}{*}{CP-violation phase}      &
   \multirow{2}{*}{$\delta_{\rm CP}$}       &
   \multirow{2}{*}{$^\circ$}                &
   \multirow{2}{*}{$\checkmark$}                &
   \multirow{2}{*}{$\checkmark$}                &
   \multirow{2}{*}{$\checkmark$}                &
   \multirow{2}{*}{230}                     &
   Present\footref{fnote2}: $230_{-36}^{+25}$  \\
   &
   &
   &
   &
   &
   &
   &
   Proj.\footref{fnote3}: Normal, $\sigma = 6.687$ \\   
  \end{tabular}
 \end{ruledtabular}  
\end{table*}
\endgroup

Within our benchmark flux scenarios (\cref{sec:scenarios}), we look for the presence of a flavor transition as a function of neutrino energy using present IceCube observations and mock projected observations by upcoming neutrino telescopes (Section~\ref{sec:detectors}).
For present data, we use the publicly available 7.5-year sample of IceCube HESE events~\cite{IceCube:2020wum, IC75yrHESEPublicDataRelease}. For projections, we simulate data with two distinct event selections, HESE and TGM, based on the PLE$\nu$M framework~\cite{Schumacher:2021hhm}, and perform our analysis with them combined. Our statistical methods are Bayesian.
We perform two main tasks: measuring the flavor composition at Earth and, based on it, inferring the flavor composition at the neutrino sources.


\subsection{Measuring the flavor composition at Earth}
\label{sec:analysis_flavor-earth}

For a given choice of spectrum shape from among our benchmarks (Section~\ref{sec:scenarios_spectrum}), our goal is to measure the flavor composition at Earth at low and high energy, $f_{\alpha, \oplus}^{\rm LE}$ and $f_{\alpha, \oplus}^{\rm HE}$, respectively.  We do so by contrasting samples of observed events, real or mock, against test event samples computed under different assumptions of spectrum shape and flavor composition, in search for the ones that describe the observations best.

Table~\ref{tab:parameters} shows the free model parameters that we vary in our analysis, and the prior distributions that we use for them later in our statistical analysis.  Which spectrum shape parameters are varied depends on our choice of spectrum shape from our PL, BPL, and Step benchmarks.  The flavor composition at LE, given by $f_{e, \oplus}^{\rm LE}$ and $f_{\mu, \oplus}^{\rm LE}$, is varied independently from the flavor composition at HE, $f_{e, \oplus}^{\rm HE}$ and $f_{\mu, \oplus}^{\rm HE}$.  (Because $\sum_\alpha f_{\alpha, \oplus}^{\rm LE} = \sum_\alpha f_{\alpha, \oplus}^{\rm HE}=1$, we only need vary the electron and muon flavor content.)

To ensure that the flavor fractions are sampled uniformly as part of our statistical analysis, we parametrize them as $(f_{e,\oplus}^{\rm LE}, f_{\mu,\oplus}^{\rm LE}, f_{\tau,\oplus}^{\rm LE}) \equiv (\sin_\oplus^2\theta^{\rm LE}  \cos^2\psi_\oplus^{\rm LE}, \sin^2\theta_\oplus^{\rm LE}  \sin^2\psi_\oplus^{\rm LE}, \cos^2\theta_\oplus^{\rm LE})$, where $\theta_\oplus^{\rm LE}$ and $\psi_\oplus^{\rm LE}$ are ancillary angles, and similarly for HE.  Then we build priors based on the Haar measure~\cite{Haba:2000be} of the volume element $df_{e,\oplus}^{\rm LE} \wedge df_{\mu,\oplus}^{\rm LE} \wedge df_{\tau,\oplus}^{\rm LE} = {\rm d}\left(\sin^4 \theta_\oplus^{\rm LE} \right) \wedge {\rm d}\left(\cos 2\psi_\oplus^{\rm LE}\right)$.  By using uniform priors to sample $\sin^4 \theta_\oplus^{\rm LE} \in \left[0,1\right]$ and $\cos 2\psi_\oplus^{\rm LE} \in \left[-1,1\right]$, and similarly for HE, we ensure an unbiased sampling of the flavor fractions.

In describing our statistical methods below, we collect the spectrum shape parameters and flavor-composition parameters into the parameter set $\mathbf{\Theta}_\oplus$, whose contents depend on the choice of benchmark flux:
\begin{description}
 \item[Single power law spectrum (PL)]
  \begin{eqnarray}
   \mathbf{\Theta}_\oplus^{\rm PL}
   &=&
   \left(
   \Phi_{\nu, 0},
   \gamma, 
   \sin^4 \theta_{\oplus}^{\rm LE}, 
   \cos 2\psi_{\oplus}^{\rm LE},
   \sin^4 \theta_{\oplus}^{\rm HE}, 
   \right.
   \nonumber
   \\
   && \qquad \left. 
   \cos 2\psi_{\oplus}^{\rm HE},
   E_\mathrm{trans}
   \right)
  \end{eqnarray}
 \item[Broken power law spectrum (BPL)]
  \begin{eqnarray}
   \mathbf{\Theta}_\oplus^{\rm BPL}
   &=&
   \left( 
   \Phi_{\nu,0}^{\rm LE},
   \Phi_{\nu,0}^{\rm HE},
   \gamma,
   \sin^4 \theta_{\oplus}^{\rm LE}, 
   \cos 2\psi_{\oplus}^{\rm LE},
   \right.
   \nonumber \\
   && \qquad \left.
   \sin^4 \theta_{\oplus}^{\rm HE}, 
   \cos 2\psi_{\oplus}^{\rm HE},
   E_\mathrm{trans}
   \right)
  \end{eqnarray}
 \item[Step-like spectrum (Step)]
  \begin{eqnarray}
   \mathbf{\Theta}_\oplus^{\rm Step}
   &=&
   \left( 
   \Phi_{\nu,0},
   \gamma^{\rm LE},
   \gamma^{\rm HE},
   \sin^4 \theta_{\oplus}^{\rm LE}, 
   \cos 2\psi_{\oplus}^{\rm LE},
   \right.
   \nonumber \\
   && \qquad \left.
   \sin^4 \theta_{\oplus}^{\rm HE}, 
   \cos 2\psi_{\oplus}^{\rm HE},
   E_\mathrm{trans}
   \right)
  \end{eqnarray}
\end{description}
Table~\ref{tab:parameters} summarizes the above parameter sets.
 

\subsection{Inferring the flavor composition at the sources}
\label{sec:analysis_flavor-sources}

Using the methods introduced in \Refe~\cite{Bustamante:2019sdb} (see also \Refe~\cite{Song:2020nfh}), we can infer the flavor composition at the neutrino sources.  The process involves undoing the effect of flavor oscillations during neutrino propagation to Earth, given the flavor composition measured at Earth and knowledge of the allowed ranges of the neutrino mixing parameters.  For the first time, we do this separately at low and high energy, \ie, we infer $f_{\alpha, {\rm S}}^{\rm LE}$ and $f_{\alpha, {\rm S}}^{\rm HE}$.  Because we use the diffuse neutrino flux, made up of the contributions from the full population of neutrino sources, what we infer is the population-averaged flavor composition at the sources.

Table~\ref{tab:parameters} show the free parameters and priors that we vary in our statistical analysis.  Because standard scenarios of high-energy astrophysical neutrino production do not yield $\nu_\tau$, when inferring the flavor composition at the sources we assume $f_{\tau, {\rm S}}^{\rm LE} = f_{\tau, {\rm S}}^{\rm HE} = 0$ and so we need only infer the electron flavor fraction at LE, $f_{e, {\rm S}}^{\rm LE}$, and at HE, $f_{e, {\rm S}}^{\rm HE}$.  For the neutrino mixing parameters, under the requirement of Haar measure, we choose parameters $\mathbf{\Omega} = \left( \sin^2 \theta_{12}, \cos^4 \theta_{13}, \sin^2 \theta_{23}, \delta_{\rm CP} \right)$, and use the same present-day and projected probability distributions (Table~\ref{tab:parameters}) that we use to compute the allowed flavor regions at Earth in Figs.~\ref{fig:triangle_main} and \ref{fig:triangles}.

In describing our statistical methods below, we collect the spectrum shape parameters and flavor-composition parameters into the parameter set $\mathbf{\Theta}_{\rm S}$, whose contents depend on the choice of benchmark flux:
\begin{description}
 \item[Single power law spectrum (PL)]
  \begin{eqnarray}
   \mathbf{\Theta}_{\rm S}^{\rm PL}
   &=&
   \left(
   \Phi_{\nu, 0},
   \gamma, 
   f_{e, {\rm S}}^{\rm LE},
   f_{e, {\rm S}}^{\rm HE},
   E_\mathrm{trans}
   \right)
   \cup
   \mathbf{\Omega}
  \end{eqnarray}
 \item[Broken power law spectrum (BPL)]
  \begin{eqnarray}
   \mathbf{\Theta}_{\rm S}^{\rm BPL}
   &=&
   \left( 
   \Phi_{\nu,0}^{\rm LE},
   \Phi_{\nu,0}^{\rm HE},
   \gamma,
   f_{e, {\rm S}}^{\rm LE},
   f_{e, {\rm S}}^{\rm HE},   
   E_\mathrm{trans}
   \right)
   \cup
   \mathbf{\Omega}
  \end{eqnarray}
 \item[Step-like spectrum (Step)]
  \begin{eqnarray}
   \mathbf{\Theta}_{\rm S}^{\rm Step}
   &=&
   \left( 
   \Phi_{\nu,0},
   \gamma^{\rm LE},
   \gamma^{\rm HE},
   f_{e, {\rm S}}^{\rm LE},
   f_{e, {\rm S}}^{\rm HE},   
   E_\mathrm{trans}
   \right)
   \cup
   \mathbf{\Omega}
  \end{eqnarray}
\end{description}
Table~\ref{tab:parameters} summarizes the above parameter sets.


\subsection{HESE likelihood}
\label{sec:analysis_hese}

The HESE MC sample~\cite{IC75yrHESEPublicDataRelease} provides, for each simulated event, primary quantities, \ie, neutrino energy, zenith angle, and flavor, and reconstructed quantities, \ie, deposited energy, reconstructed  zenith angle, morphology ---cascade (c), track (tr), or double cascade (dc)---and, for double cascades, reconstructed track length.  As in the analysis of the 7.5-year HESE sample by the IceCube Collaboration~\cite{IceCube:2020wum}, we use HESE events detected with energies above 60~TeV.

In searches for astrophysical neutrinos, the background is due to atmospheric muons and neutrinos.  In our analysis, we fix their fluxes to their best-fit values found by IceCube, in Table VI.1 of \Refe~\cite{IceCube:2020wum}. Since the atmospheric contribution is subdominant above $60$~TeV and negligible above $100$~TeV, this choice does not significantly impacts our results. We also fix the parameters governing detector systematic uncertainties, including the light acceptance of the optical module and ice properties, to their best-fit values reported in \Refe~\cite{IceCube:2020wum}. 

In the MC sample, the $j$-th event has an associated weight $w_j$ that depends only on $\mathbf{\Theta}_\oplus$ when measuring the flavor composition at Earth (Section~\ref{sec:analysis_flavor-earth}), or only on $\mathbf{\Theta}_{\rm S}$ when inferring the flavor composition at the sources (Section~\ref{sec:analysis_flavor-sources}). The weights allow us transform the HESE MC sample and generate Asimov event samples for different choices of the spectrum and flavor composition of the astrophysical neutrinos.
For cascades and tracks, we compute the number of events in bins of deposited energy and reconstructed zenith angle; for double cascades, we compute it in bins of deposited energy and track length.  We use the same binning scheme as in the IceCube HESE analysis, from Table V.1 of \Refe~\cite{IceCube:2020wum}.  

Using the weights, we compute the expected number of events of each morphology, m = \{c, tr, dc\}, due to astrophysical neutrinos.  In the $i$-th bin, this is
\begin{equation}
 \mu_{{\rm m}, i}^{\rm ast}(\mathbf{\Theta})
 =
 \sum_{j\in {\rm bin}~i} w_{{\rm m}, j}(\mathbf{\Theta}) \;,
\end{equation}
where the sum is over the events that fall in that bin.  The total number of expected events, $\mu_{{\rm m}, i}$,  includes also the contribution from atmospheric backgrounds, $\mu_{{\rm m}, i}^{\rm atm}$, \ie,
\begin{equation}
 \label{equ:rate_mean}
 \mu_{{\rm m}, i}(\mathbf{\Theta})
 =
 \mu_{{\rm m}, i}^{\rm ast}(\mathbf{\Theta})
 +
 \mu_{{\rm m}, i}^{\rm atm} \;.
\end{equation}
We also compute the expected variance,
\begin{equation}
 \sigma^2_{{\rm m}, i}(\mathbf{\Theta})
 =
 \sum_{j\in {\rm bin}~i} w_{{\rm m}, j}^2(\mathbf{\Theta}) 
 +
 (\mu_{{\rm m}, i}^{\rm atm})^2
 \;,
\end{equation}
to account for statistical fluctuations in event rate.

The expected event numbers and fluctuations are then compared against a sample of observed events, real for present-day data and mock for projections, which contains $N_{{\rm m}, i}$ events of morphology m in the $i$-th bin.  For present-day results, we use the IceCube 7.5-year HESE sample~\cite{IceCube:2020wum, IC75yrHESEPublicDataRelease}.  For projections, we use Asimov data samples built with the mean expected rate, \equ{rate_mean}, \ie, $N_{{\rm m}, i} = \mu_{{\rm m}, i}$.
The comparison is performed using the same effective likelihood function used by the IceCube Collaboration in \Refe~\cite{IceCube:2020wum}, \ie,
\begin{widetext}
\begin{equation}
 \mathcal{L}_{\mathrm{HESE}}(\mathbf{\Theta})
 =
 \prod_{\rm m}^{\{{\rm c}, {\rm tr}, {\rm dc}\}}
 \prod_i^{\rm bins} 
 \left(\frac{\mu_{{\rm m}, i}}{\sigma_{{\rm m}, i}^2}\right)^{\frac{\mu_{{\rm m}, i}^2}{\sigma_{{\rm m}, i}^2}+1} 
 \Gamma\left(N_{{\rm m}, i}+\frac{\mu_{{\rm m}, i}^2}{\sigma_{{\rm m}, i}^2}+1\right)
 \left[N_{{\rm m}, i}!
 \left(1+\frac{\mu_{{\rm m}, i}}{\sigma_{{\rm m}, i}^2}\right)^{N_{{\rm m}, i}+\frac{\mu_{{\rm m}, i}^2}{\sigma_{{\rm m}, i}^2}+1} \Gamma\left(\frac{\mu_{{\rm m}, i}^2}{\sigma_{{\rm m}, i}^2}+1\right)\right]^{-1} \;,
 \label{equ:likelihood}    
\end{equation}    
\end{widetext}
where $\Gamma$ is the gamma function. This likelihood function, introduced in \Refe~\cite{Arguelles:2019izp}, takes into account the statistical fluctuations of the MC-based predictions and of the observed data, preventing biased results and providing improved parameter coverage.


\subsection{TGM likelihood}
\label{sec:analysis_tgm}

As mentioned earlier (Section~\ref{sec:detectors_event-types_hese-tgm}), using TGM events together with HESE events tightens the measurement of the muon-flavor content. However, unlike HESE, there is no publicly available present-day sample of the TGM detected by IceCube~\cite{IceCube:2021uhz} that we can use to compute a likelihood to the same level of detail as we do for HESE. Therefore, we do not include TGM in our present-day results, but we do when making projections of future experimental sensitivity, similarly to \Refes~\cite{IceCube:2015gsk, Naab:2023xcz}.

We compute the expected number of TGM events in the $i$-th bin of reconstructed muon energy, $\mu_{{\rm TGM}, i}$, by convolving the fluxes of atmospheric plus astrophysical neutrinos, the latter of which depends on the model parameters, $\mathbf{\Theta}_\oplus$ or $\mathbf{\Theta}_{\rm S}$, with the effective detector area for the detection of TGM, and add the subdominant contribution of atmospheric muons. TGM events are dominated by atmospheric neutrinos, especially at LE; we fix their flux to the computation by \texttt{MCEq} with cosmic-ray flux model H3a, hadronic model SIBYLL-2.3c and atmosphere model NRLMSISE-00~\cite{MCEq, picone2002nrlmsise, Gaisser:2011klf, Riehn:2017mfm}.

We compute TGM event rates using the PLE$\nu$M framework~\cite{Schumacher:2021hhm}, which estimates the effective area of future neutrino telescopes as a function of declination by rotating the IceCube effective area to the detector position and
scaling it by the detector size. The IceCube effective area is based on a recent public IceCube point-source data release~\cite{IceCube:2021xar}. The computation of event rates includes the same energy smearing provided in the data release, to account for event reconstruction effects. Because of the difficulty in modeling the contamination of atmospheric muons in the sample, which dominate for down-going events, we only consider up-going events, where the contamination is small.  The event selection of the IceCube point-source sample is different from the TGM event selection used for diffuse flux measurements in, \eg, \Refe~\cite{IceCube:2021uhz}, and contains a higher atmospheric background.  However, since we do not use the point-source data directly in our work, but only its associated detector effective area determined by simulation for an up-going neutrino flux, we do not expect a notable difference between basing our results on it {\it vs.}~basing them on the alternative TGM event selection that is used for diffuse flux measurements. 

The way we generate our samples of TGM results in them including also HESE tracks, \ie, tracks that start inside the detector volume, which can lead to overcounting when combined with HESE data.  However, this is not significant, since the number of HESE tracks is far smaller than the number of TGM; see \Cref{sec:detectors_event-types-tgm}.

We include the contribution of TGM via the Poisson likelihood
\begin{equation}
 \mathcal{L}_{\mathrm{TGM}} \left(\mathbf{\Theta}\right)
 =
 e^{-\mu_{\rm TGM}\left(\mathbf{\Theta}\right)}
 \prod_i^{\rm bins} 
 \frac{\left( N_{{\rm TGM}, i} \right)^{-\mu_{{\rm TGM}, i}\left(\mathbf{\Theta}\right)}}
 {N_{{\rm TGM}, i}!} \;,
\end{equation}
where the product is over bins of reconstructed muon energy, $E_\mu$, each of size 0.2 in $\log_{10}(E_\mu/{\rm GeV})$, $\mu_{\rm TGM} \equiv \sum_i \mu_{{\rm TGM}, i}$ is the total number of events across all bins, and $N_{{\rm TGM}, i}$ is the number of observed tracks in the projected mock event sample.   Unlike the HESE sample, where the events have a rather high energy threshold of 60~TeV, we count TGM with energies $E_\mu \gtrsim 100$~GeV.


\subsection{Combined HESE and TGM likelihood}
\label{sec:analysis_hese-tgm}

In our projections, for a choice of astrophysical neutrino spectrum shape and flavor composition from among our benchmarks, we generate mock samples of  HESE events, $N_{{\rm c}, i}$, $N_{{\rm tr}, i}$, $N_{{\rm dc}, i}$, and of TGM events, $N_{{\rm TGM}, i}$.  With them, we compute the combined likelihood
\begin{equation}
 \mathcal{L}(\mathbf{\Theta})
 =
 \mathcal{L}_{\mathrm{TGM}}(\mathbf{\Theta})
 \mathcal{L}_{\mathrm{HESE}}(\mathbf{\Theta}) \;.
\end{equation}


\subsection{Posterior and parameter optimization}

Based on the above likelihood functions, we compute the joint parameter posterior, 
\begin{equation}
 \label{equ:posterior}
 \mathcal{P}(\mathbf{\Theta})
 =
 \mathcal{L}(\mathbf{\Theta})
 \pi(\mathbf{\Theta}) \;,
\end{equation}
where $\pi(\mathbf{\Theta})$ is the joint prior distribution of the model parameters.  Table~\ref{tab:parameters} shows our choices of priors.  We maximize the posterior numerically, using Markov Chain Monte Carlo methods implemented in \texttt{UltraNest}~\cite{Buchner:2014, Buchner:2017, Ultranest}, an efficient Bayesian importance sampler.  Later, we report two-dimensional credible intervals of the flavor content by marginalizing over all other parameters (\figu{triangles}).  In Appendix~\ref{app:posteriors}, we also report
one-dimensional intervals for selected parameters.


\section{Results}
\label{sec:results}

\begin{figure*}[htb!]
 \centering
 \subfigure{%
  \includegraphics[width=0.38\linewidth]{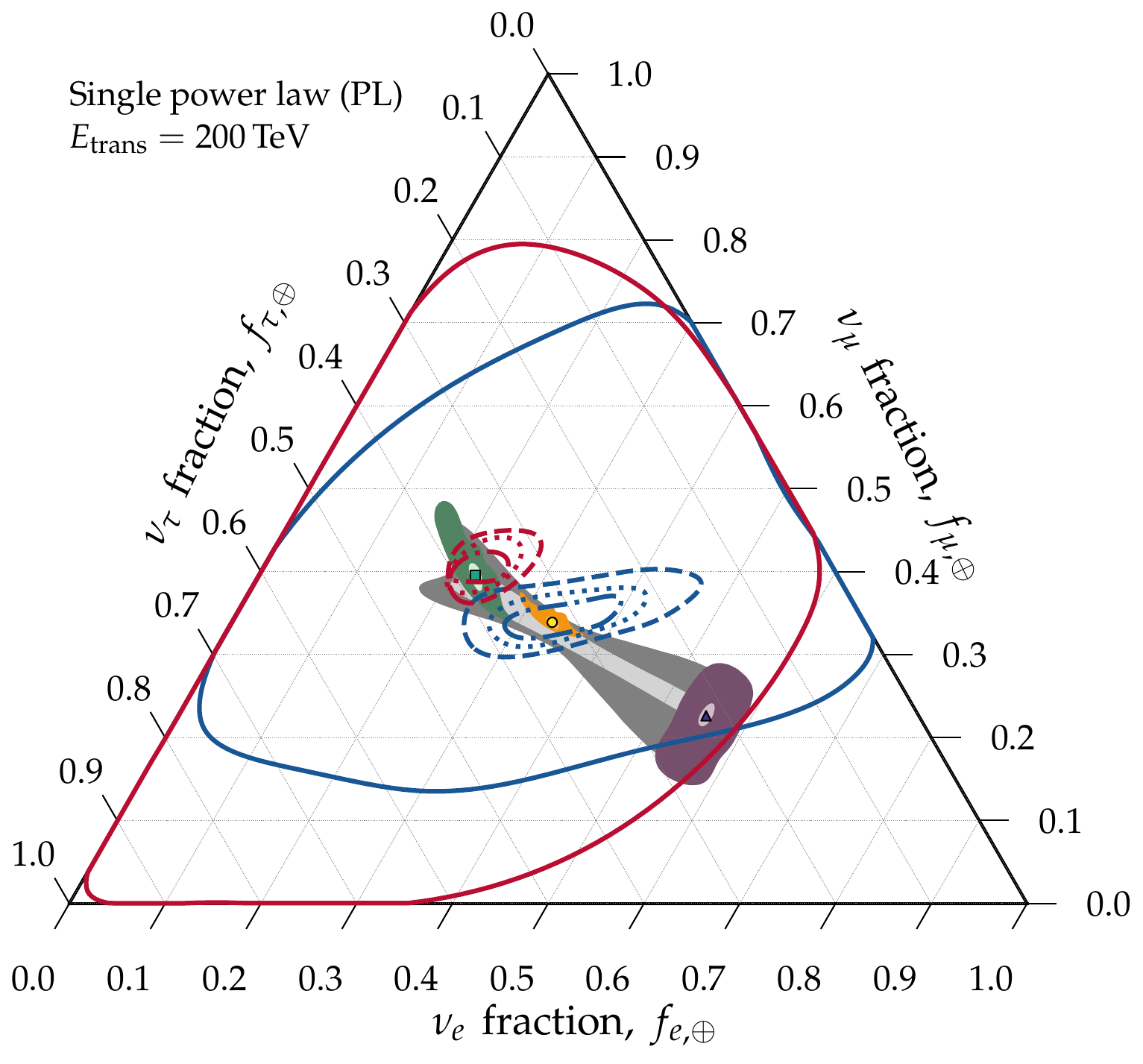}
 }
 \subfigure{
  \includegraphics[width=0.38\linewidth]{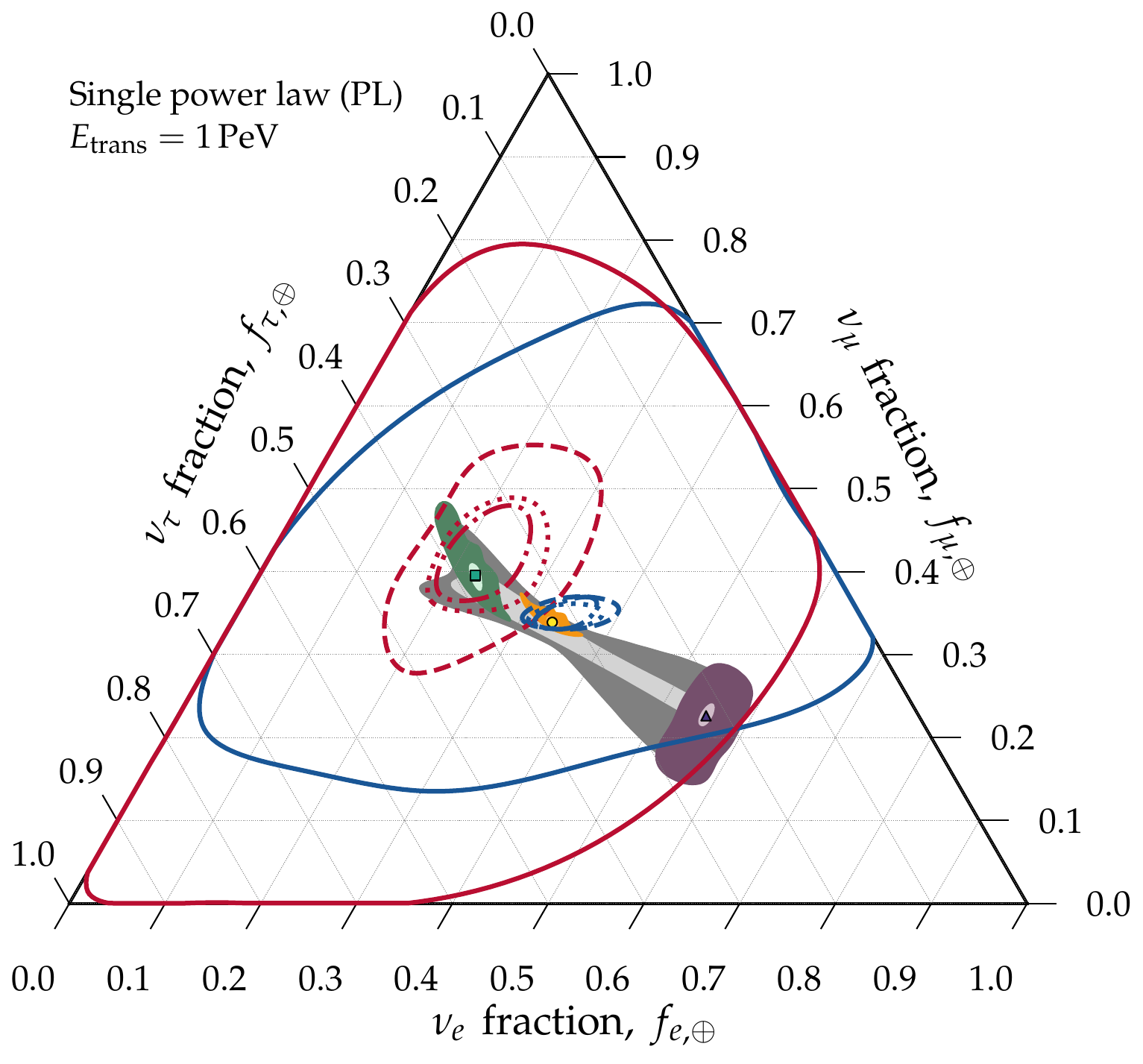}
 }\\ 
 \vspace*{-2.5mm}
 \subfigure{
  \includegraphics[width=0.38\linewidth]{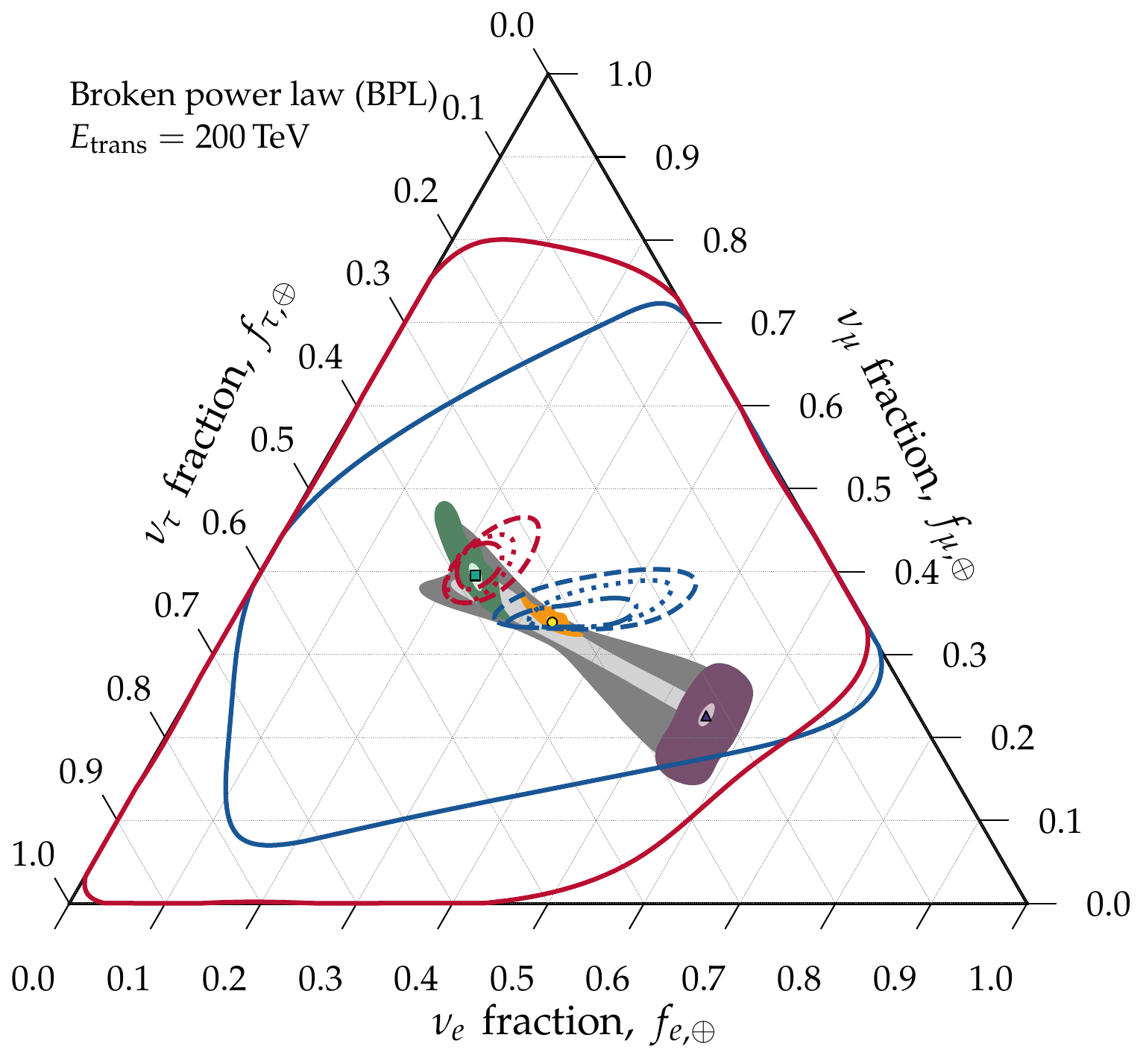}
 }    
 \subfigure{
  \includegraphics[width=0.38\linewidth]{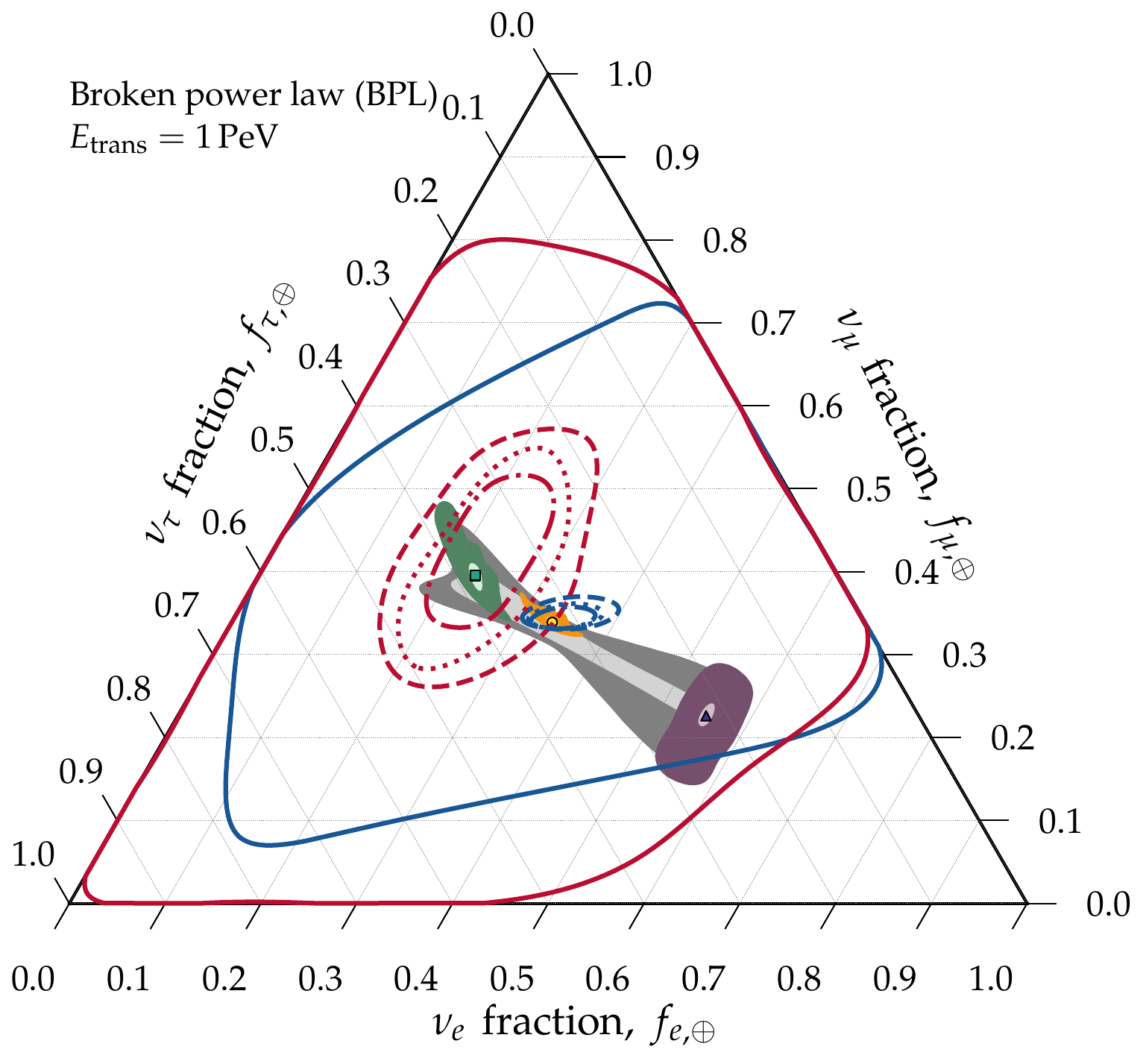}
 }\\
 \vspace*{-2.5mm}
 \subfigure{
  \includegraphics[width=0.38\linewidth]{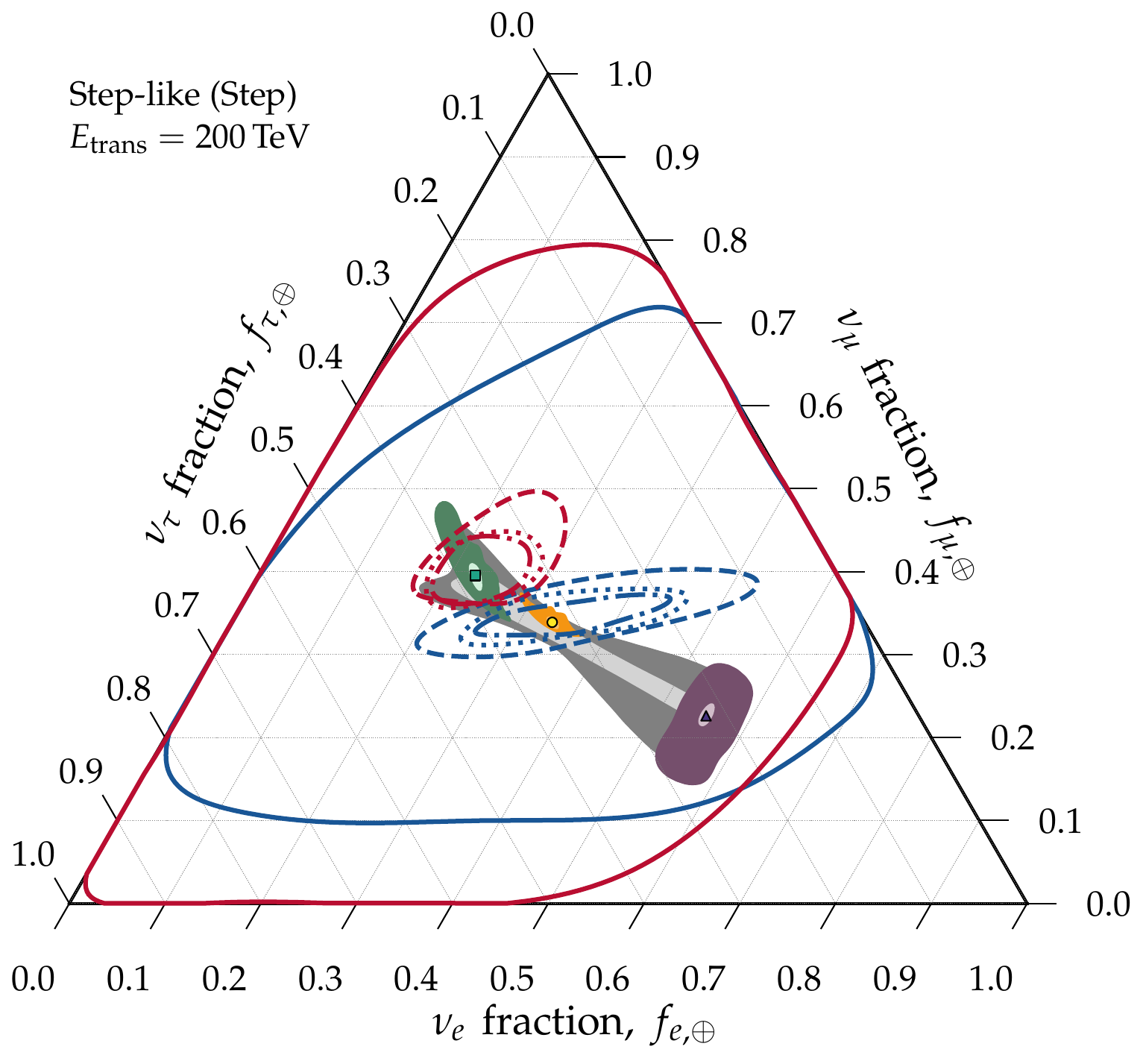}
 }    
 \subfigure{
  \includegraphics[width=0.38\linewidth]{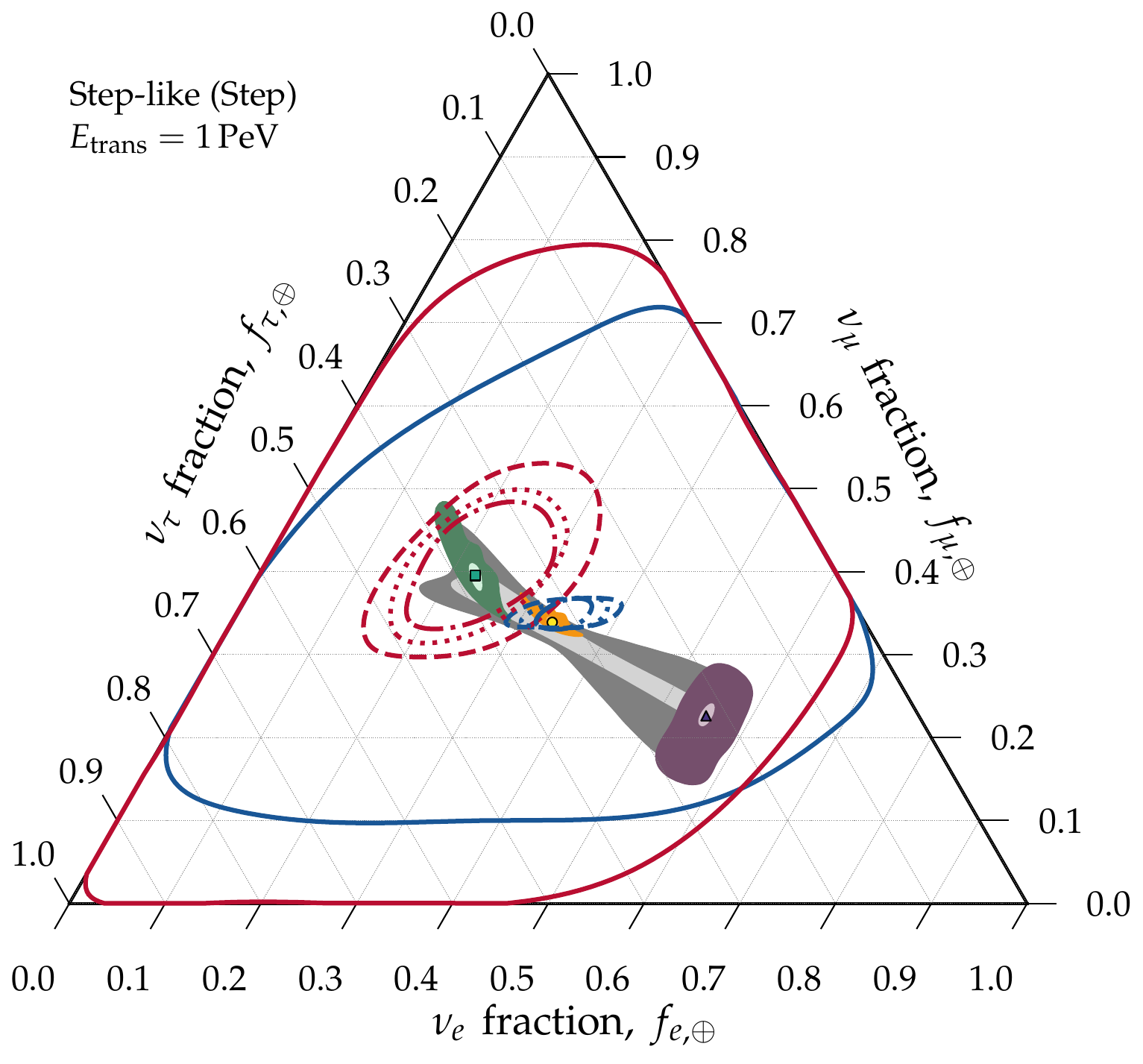}
 }\\
 \vspace*{-2mm}
 \includegraphics[width=\linewidth]{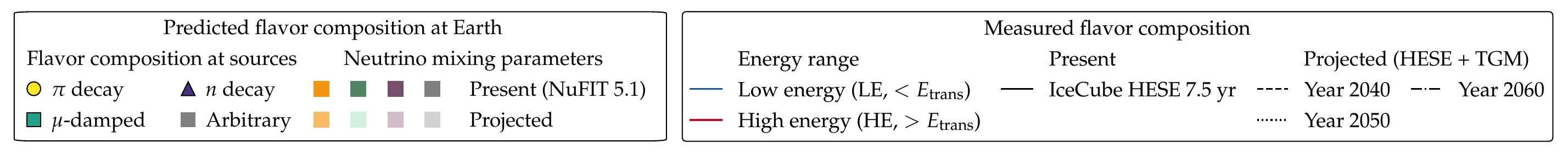}
 \vspace*{-7mm}
 \caption{\textbf{\textit{Flavor composition of high-energy astrophysical neutrinos measured at Earth, at LE and HE.}}  Present measurements are from the IceCube 7.5-year HESE sample~\cite{IceCube:2020wum, IC75yrHESEPublicDataRelease}; projected measurements, from the detection of HESE plus TGM by multiple neutrino telescopes (Table~\ref{tab:detectors}).  Results are for three benchmark neutrino spectra (\figu{benchmark_flux})---PL (\textit{top}), BPL (\textit{center}), and Step (\textit{bottom})---that transition at $E_\textrm{trans} = 200$~TeV (\textit{left}) or 1~PeV (\textit{right}).  For projections, the transition is from pion decay at LE to muon-damped at HE. The predicted regions of flavor composition at Earth are from standard oscillations~\cite{Bustamante:2015waa, Song:2020nfh} Table~\ref{tab:parameters}).  All regions are at 68\%~C.L. See \Cref{sec:results_flavor-earth} for details.  \textit{There is no evidence of a flavor transition with energy in present HESE data, but projections reveal sensitivity to detect one.}}
 \label{fig:triangles}
\end{figure*}

First (\Cref{sec:results_flavor-earth}), we show measurements of the LE and HE flavor composition at Earth, for each of our benchmark scenarios of neutrino spectrum, using present-day IceCube HESE events and projected event samples detected by future neutrino telescopes. 
Then (\Cref{sec:results_model-comparison}), we quantify the power to infer the presence of a flavor transition via Bayes factors that contrast the evidence for descriptions of the observed data with and without a flavor transition. 
Finally (\Cref{sec:results_flavor-sources}), we infer the flavor composition at the sources for each of our benchmark scenarios.


\subsection{Flavor composition at Earth}
\label{sec:results_flavor-earth}

We compute the posterior distributions corresponding to the flavor composition at Earth using the methods outlined in~\Cref{sec:analysis}, separately for each of our three benchmark neutrino spectra: PL, BPL, and Step.  For our present-day results, we use the IceCube 7.5-year HESE sample~\cite{IceCube:2020wum, IC75yrHESEPublicDataRelease}.  For our projections, we assume that the true flavor composition is that of pion decay below the transition energy, $E_\textrm{trans}$, and muon-damped above it; we show results assuming $E_\textrm{trans} = 200$~TeV and 1~PeV.  We extract the flavor composition at LE (below $E_\textrm{trans}$) and HE (above $E_\textrm{trans}$) using the methods presented in \Cref{sec:analysis}.

Figure~\ref{fig:triangles} shows the resulting posterior distributions of the flavor composition.   At present, due to the limited size of the HESE sample, the 68\%~credible regions cover nearly the entire flavor triangle regardless of the choice of benchmark spectrum.  The measured transition energy assuming the PL flux is $E_\textrm{trans} \approx 193_{-87}^{+2777}$~TeV.  The large errors render the measurement meaningless; we find similar results  the BPL and Step fluxes (see Table~\ref{tab:results_earth} in Appendix~\ref{app:posteriors}).  The large errors reflect the fact that there is no statistically significant evidence for a flavor transition in present-day HESE data; we quantify later (\Cref{sec:results_model-comparison}). 
Appendix~\ref{app:posteriors} shows the joint posteriors and one-dimensional marginalized allowed ranges of the free model parameters, including the flavor composition. 

\begin{figure}[t!]
 \centering
 \includegraphics[width=\columnwidth]{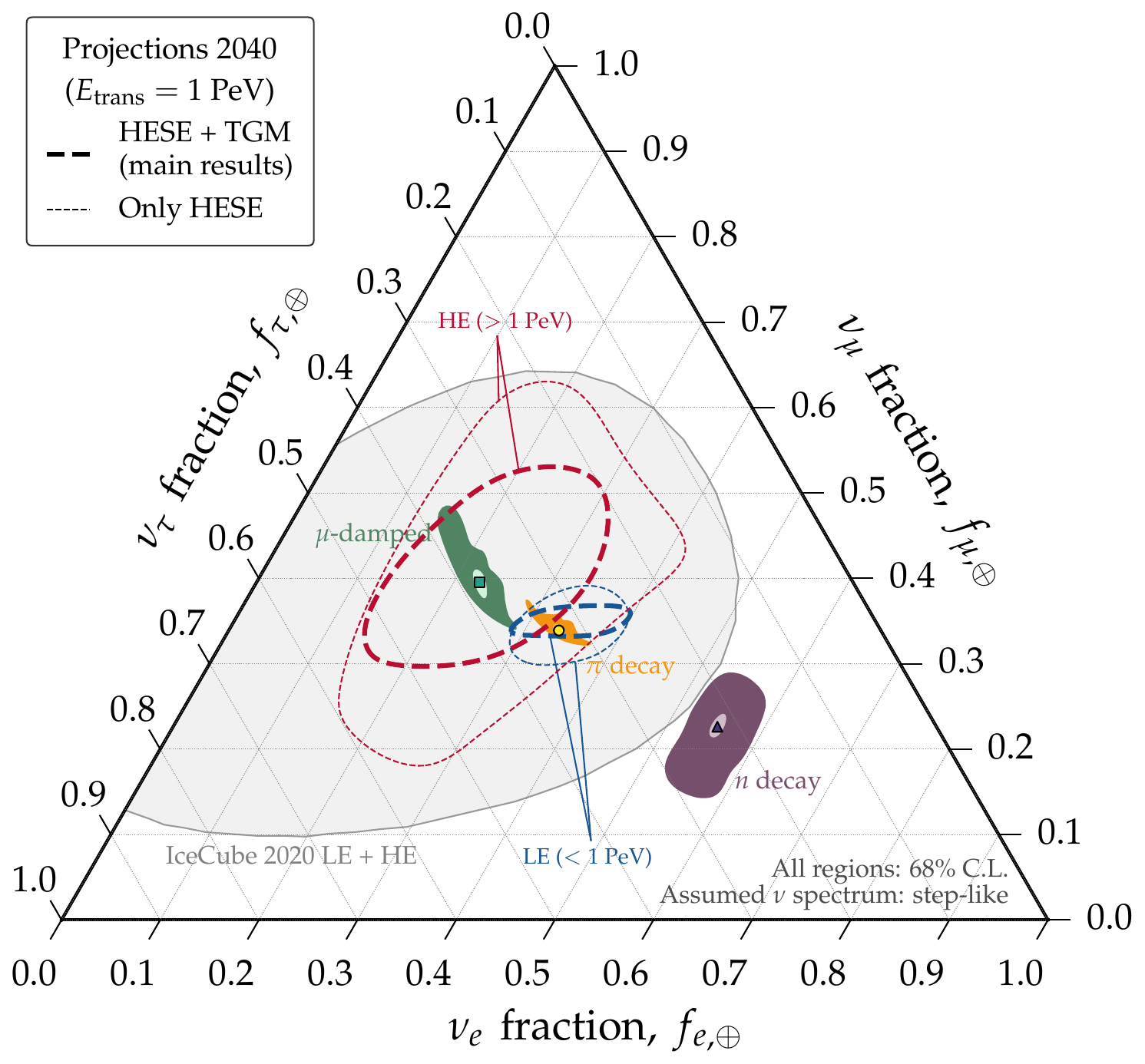}
 \caption{\textbf{\textit{Improvement in flavor-composition measurements due to TGM.}}  This plot shows projections for the year 2040.  Our main results are generated using HESE plus TGM; for comparison, we show results generated using only HESE.  The neutrino spectrum is assumed to be step-like (\figu{benchmark_flux}). Previous energy-averaged measurements are from IceCube (``IceCube 2020 LE + HE)~\cite{IceCube:2020fpi}.  The allowed flavor regions for neutrino production via pion decay, muon-damped production, and neutron decay are computed as in \Refe~\cite{Song:2020nfh}, assuming present-day~\cite{Esteban:2020cvm, nufit5.1} and projected~\cite{Song:2020nfh} uncertainties in the neutrino mixing parameters.  See \Cref{sec:results_flavor-earth} and \figu{step_corner} in Appendix~\ref{app:posteriors} for details.  \textit{Using TGM improves the measurement of the $\nu_\mu$ content roughly by a factor of 2.}}
 \label{fig:triangle_hese_vs_tgm}
\end{figure}

Figure~\ref{fig:triangles} also shows important improvement in the projected measurements in the milestone years 2040, 2050, and 2060, tough results for the latter two milestones should be taken to be especially tentative.  The improvement in between today and 2040 is vast: the uncertainty at the 68\%~C.L. shrinks by a factor of 5--6 for both LE and HE.  After 2040, the improvement slows down because the fractional increase in the combined detector exposure over time is smaller and because the measurement becomes increasingly marred by systematic uncertainties such as the event reconstruction capabilities of the detectors, whose improvement over time, while likely, is not factored into our projections.

\begin{figure}[ht!]
 \centering
 \hspace*{0.5cm}
 \includegraphics[scale=0.42]{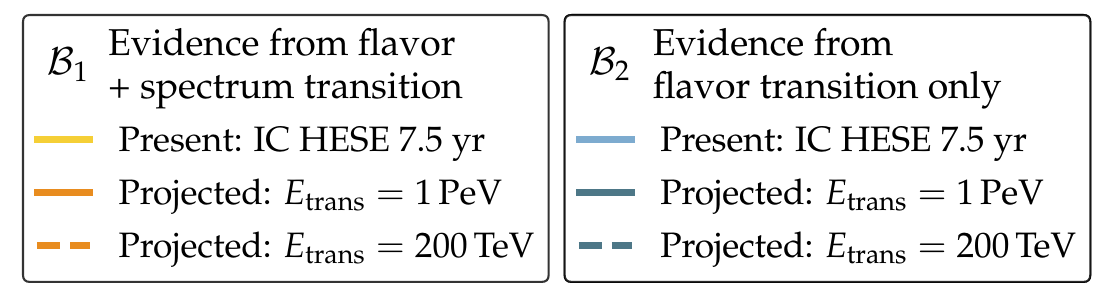}
 \includegraphics[scale=0.385]{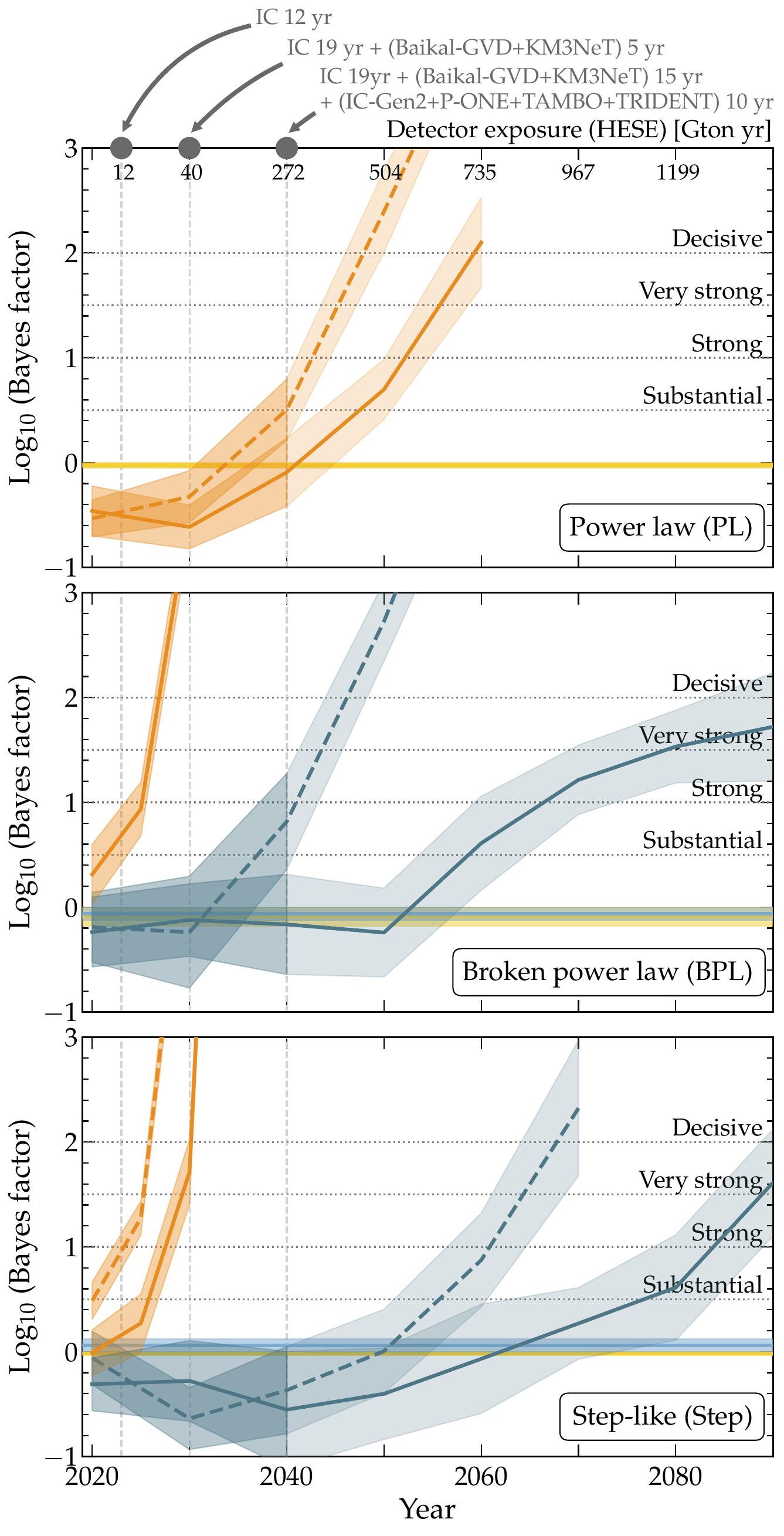}
 \vspace*{-0.75cm}
 \caption{
 \textbf{\textit{Evidence for a transition with energy in the flavor composition of high-energy astrophysical neutrinos.}} 
 Results show preferred and 68\%~C.L.~allowed values of Bayes factors $\mathcal{B}_1$ and $\mathcal{B}_2$ (Table~\ref{tab:bayes}).  Present results are from the IceCube 7.5-year HESE sample~\cite{IceCube:2020wum, IC75yrHESEPublicDataRelease}; projected results, from HESE plus TGM detected by multiple neutrino telescopes (Table~\ref{tab:detectors}).  Results are for three benchmark neutrino spectra (\figu{benchmark_flux})---PL (\textit{top}), BPL (\textit{center}), and Step (\textit{bottom})---that transition at $E_\textrm{trans}$.  Beyond 2040, projections are especially tentative.  
 See \Cref{sec:results_model-comparison} for details.
 \textit{There is no evidence of a flavor transition in present HESE data, but it may be decisive by 2030.} 
 \vspace*{-1.0cm}}
 \label{fig:bfactor}
\end{figure}

Figure~\ref{fig:triangle_hese_vs_tgm} illustrates one of the two factors behind the improvement in the projected measurements: including though-going muons shrinks the measurement errors mainly in the $f_{\mu, \oplus}$ direction.  The second factor is the large increase in the combined exposure of upcoming detectors (\figu{bfactor}) to HESE events, which shrinks the measurement errors mainly in the $f_{e, \oplus}$ direction, as evidenced by comparing Figs.~\ref{fig:triangle_hese_vs_tgm} and \ref{fig:triangles}.

The key to achieving the sensitivity to a flavor transition is the value of the transition energy, which determines the number of events from which to measure the flavor composition in the LE and HE regions.  If the flavor transition were to happen at too low or too high an energy, the number of events in the LE or HE region would not be large enough to pin down the flavor composition in that region, thus curbing attempts to identify the transition.  Therefore, neutrino telescopes are most sensitive to an intermediate transition energy.  

Nevertheless, \figu{triangles} shows that future measurements should be able to distinguish between the LE and HE flavor composition, for both of our choices $E_\textrm{trans}$, 200~TeV and 1~PeV, and regardless of the flux model used. The sizes of the 68\%~C.L.~allowed regions of flavor composition are comparable for all flux models.  By 2040, the distinction is marginal at 68\%~C.L.; by 2050, it reaches 95\%~C.L. (not shown).  Given that the neutrino flux falls steeply with energy, a value of $E_\textrm{trans}$ of 200~TeV splits the event samples more evenly between LE and HE than a value of 1~PeV.  As a result, \figu{triangles} shows that, for $E_\textrm{trans} = 200$~TeV, the measurements at LE and HE have comparable uncertainty, with HE marginally worse than LE, while for $E_\textrm{trans} = 1$~PeV the measurement at HE is about a factor-of-3 more uncertain than at LE, due to the paucity of events in the HE region.

Although the measurement uncertainty of the flavor composition is comparable for the PL, BPL, and Step flux models, the precision with which $E_\textrm{trans}$ is measured is different in each of them.  For instance, assuming 200~TeV as the true value of $E_\textrm{trans}$, its best-fit measured value is measured accurately between 191--199~TeV, depending on the flux model, but the 68\%~C.L.~precision is about 53\% for PL, 9\% for BPL, and 5\% for Step.  The precision obtained when assuming a true value of 1~PeV is similar; Table~\ref{tab:results_earth} shows the full results.  

This hierarchy in precision among flux models is unsurprising: in the PL model, any evidence of flavor transition stems solely from the change in the flavor composition, while in the BPL and Step models, it comes also, and dominantly, from the change in the shape of the spectrum at $E_\textrm{trans}$ (\figu{benchmark_flux}). The precision is highest for the Step model because the change in the spectrum is the largest among our benchmark models; this is most evident when the transition is at 1~PeV (Table~\ref{tab:results_sources}).  It is auspicious that the Step model is closest to what a realistic flavor transition from pion decay to muon-damped might look like.

\subsection{Quantifying the evidence for a flavor transition}
\label{sec:results_model-comparison}

\begin{table*}[thb!]
 \begin{ruledtabular}  
  \caption{\label{tab:bayes}\textbf{\textit{Scenarios of model comparison, quantified via Bayes factors.}}  We explore two scenarios: one where evidence for a transition from LE to HE in the diffuse neutrino flux stems from flavor and spectrum changes, $\mathcal{B}_1$, and another where it stems from flavor changes only, $\mathcal{B}_2$.  In all cases, the evidence $\mathcal{Z}_{\rm B}$ is evaluated assuming no flavor transition, \ie, $f_{\alpha, \oplus}^{\rm LE} = f_{\alpha, \oplus}^{\rm HE}$ ($\alpha = e, \mu, \tau$).  We compute Bayes factors for our benchmark neutrino spectra (Section~\ref{sec:scenarios_production})---single power law (PL), broken power law (BPL), and step-like (Step).  See \figu{bfactor} for results based on present and projected data, and Section~\ref{sec:results_model-comparison} for details.}
  \centering
  \renewcommand{\arraystretch}{1.35}
  \begin{tabular}{cccc}
   \multirow{2}{*}{\makecell{Bayes factor,\\$\mathcal{Z}_{\rm A}/\mathcal{Z}_{\rm B}$}}            &
   \multirow{2}{*}{What it tests}                      &
   \multicolumn{2}{c}{Statistical evidence computed with}  \\
   \cline{3-4}
   &
   &
   $\mathcal{Z}_{\rm A}$                                     &
   $\mathcal{Z}_{\rm B}$                                     \\
   \hline
   $\mathcal{B}_1$                                     &
   \multirow{2}{*}{\makecell{Flavor and spectrum transition\\{\it vs.}~no flavor nor spectrum transition}}            &
   PL                                                  &
   PL with $f_{\alpha, \oplus}^{\rm LE} = f_{\alpha, \oplus}^{\rm HE}$                                                \\
   &
   &
   BPL                                                 &
   PL with $f_{\alpha, \oplus}^{\rm LE} = f_{\alpha, \oplus}^{\rm HE}$                                                \\
   &
   &
   Step                                                &
   PL with $f_{\alpha, \oplus}^{\rm LE} = f_{\alpha, \oplus}^{\rm HE}$                                                \\
   $\mathcal{B}_2$                                     &
   \multirow{2}{*}{\makecell{Flavor and spectrum transition\\{\it vs.}~no flavor transition}}                         &
   PL                                                  &
   PL with $f_{\alpha, \oplus}^{\rm LE} = f_{\alpha, \oplus}^{\rm HE}$                                                \\
   &
   &
   BPL                                                 &
   BPL with $f_{\alpha, \oplus}^{\rm LE} = f_{\alpha, \oplus}^{\rm HE}$                                                \\
   &
   &
   Step                                                &
   Step with $f_{\alpha, \oplus}^{\rm LE} = f_{\alpha, \oplus}^{\rm HE}$                                                \\
  \end{tabular}
 \end{ruledtabular}  
\end{table*}

Next, we quantify the preference for the presence of a flavor transition via a Bayes factor that compares the Bayesian evidence for a scenario explaining the observed events that posits a transition {\it vs.}~a scenario that does not.  The evidence is the integral of the posterior, $\mathcal{P}$ in \equ{posterior}, over the entire space of model parameter, $\mathbf{\Theta}$, \ie, $\mathcal{Z} = \int d\mathbf{\Theta} \mathcal{P}(\mathbf{\Theta})$.
The Bayes factor comparing two scenarios, A and B, respectively with evidence $\mathcal{Z}_{\rm A}$ and $\mathcal{Z}_{\rm B}$, is $\mathcal{B} \equiv \mathcal{Z}_{\rm A}/\mathcal{Z}_{\rm B}$.  The higher the value of $\mathcal{B}$, the stronger the evidence in favor of scenario A over B is in the observations.  We adopt Jeffreys' criteria~\cite{jeffreys1998theory} to interpret the values of the Bayes factor (\figu{bfactor}): $0 \leq \log_{10} \mathcal{B} < 0.5$ represents evidence for scenario A that is barely worth  mentioning; $0.5 \leq \log_{10} \mathcal{B} \leq 1$, substantial evidence; $1 \leq \log_{10} \mathcal{B} < 1.5$, strong evidence; $1.5 \leq \log_{10} \mathcal{B} < 2$, very strong evidence; and $\log_{10} \mathcal{B} \geq 2$, decisive evidence.

Table~\ref{tab:bayes} shows the two sets of Bayes factors that we compute: $\mathcal{B}_1$ and $\mathcal{B}_2$.  Via $\mathcal{B}_1$, we assess the preference for a scenario with a flavor and spectrum transition---scenario A, computed in turn for PL, BPL, and Step---\textit{vs.}~a scenario with no flavor or spectrum transition---scenario B, fixed to PL with $f_{\alpha, \oplus}^{\rm LE} = f_{\alpha, \oplus}^{\rm HE}$.  Via $\mathcal{B}_2$, we assess the preference for a scenario with a flavor and spectrum transition---scenario A, computed in turn for PL, BPL, and Step---\textit{vs.}~a scenario with no flavor transition, only spectrum transition---scenario B, matching the same choice as for scenario A but with $f_{\alpha, \oplus}^{\rm LE} = f_{\alpha, \oplus}^{\rm HE}$.
Thus, $\mathcal{B}_1$ quantifies the evidence for one of our three benchmark scenarios with flavor transition against the null hypothesis of no flavor transition and a simple power-law spectrum, whilst $\mathcal{B}_2$ asks the more difficult question of establishing the existence of a flavor transition without relying on an associated change in the spectrum shape.

Our present-day results are based on the IceCube 7.5-year HESE sample, and $\mathcal{Z}_{\rm A}$ and $\mathcal{Z}_{\rm B}$ are computed from it.  Our projections are based on the combined detection of HESE and TGM by multiple upcoming neutrino telescopes.  In the projections, we assume that the true PL, BPL, and Step fluxes are given by the parameters in Table~\ref{tab:parameters}. 
These are the fluxes with which we generate the mock event samples that we assume to be the observed ones, and $\mathcal{Z}_{\rm A}$ and $\mathcal{Z}_{\rm B}$ are computed from them.  We compute projections for the year 2040 and beyond; like before, after 2040 our projections should be understood to be especially tentative.  We use \texttt{UltraNest}~\cite{Buchner:2014,Buchner:2017} to compute the evidence by varying all free model parameters simultaneously (\Cref{sec:analysis_flavor-earth}).

Figure~\ref{fig:bfactor} shows our results.  Present-day results show no preference for a flavor transition, with or without an accompanying spectrum transition, and so $\mathcal{B}_1 \approx 1$ and $\mathcal{B}_2 \approx 1$ for all flux models.  The main limitation is the paucity of present data, which does not allow us to draw preference for any scenario.  This aligns with our present inability to measure the flavor composition at Earth (Section~\ref{sec:results_flavor-earth}) and at the sources (later, in Section~\ref{sec:results_flavor-sources}).

Fortunately, our projections in \figu{bfactor} show that the above limitation might be overcome before 2040, possibly a decade earlier, depending on the neutrino flux.  Below, we describe the features of our projections.

At any one time, $\mathcal{B}_1$ is larger than $\mathcal{B}_2$, which reflects the fact that it is easier to find preference for a transition in energy when evidence for it comes both from a change in flavor and in spectrum \textit{vs.}~when the spectrum is known and evidence comes only from a flavor change.  

The choice of transition energy has a large effect on how soon we can obtain strong evidence for a flavor transition.  We show results assuming $E_\textrm{trans} = 200$~TeV and 1~PeV.  As when measuring the flavor composition at Earth (\Cref{sec:results_flavor-earth}), a lower value of $E_\textrm{trans}$ splits the number of detected neutrinos more evenly between the low- and high-energy regions, and makes it easier to spot a transition between them. 
As a result, \figu{bfactor} shows that the Bayes factor for $E_\textrm{trans} = 200$~TeV is consistently higher than for 1~PeV.  

The preference for a transition in energy, quantified via $\mathcal{B}_1$, is higher for the BPL and Step flux models, which have a spectrum transition associated to the flavor transition, than for the PL flux model, which does not.  For the BPL and Step flux models, decisive evidence for a transition may become available already by 2030, regardless of the value $E_\textrm{trans}$, using only a subset of the future telescopes under consideration---IceCube, Baikal-GVD, and KM3NeT (Table~\ref{tab:detectors}).  For the PL flux model, substantial evidence may become available by 2040 if $E_\textrm{trans} = 200$~TeV, and decisive evidence would have to wait one or two more decades. 

Finally, \figu{bfactor} shows a subtle feature in the time evolution of our projections: the Bayes factor drops at early times and grows at later times.  This is because at early times $\mathcal{Z}_{\rm A}$ decreases as a result, first, of splitting up the limited number of detected events between the LE and HE regions and, second, of the fact that $\mathcal{Z}_{\rm A}$ depends on more free parameters than $\mathcal{Z}_{\rm B}$, for which $f_{\alpha, \oplus}^{\rm LE} = f_{\alpha, \oplus}^{\rm HE}$ by construction.  At later times, the Bayes factor instead grows, since there are enough events in the LE and HE regions for the transition between them to be identified.

\begin{figure}[t!]
 \centering
 \includegraphics[width=\columnwidth]{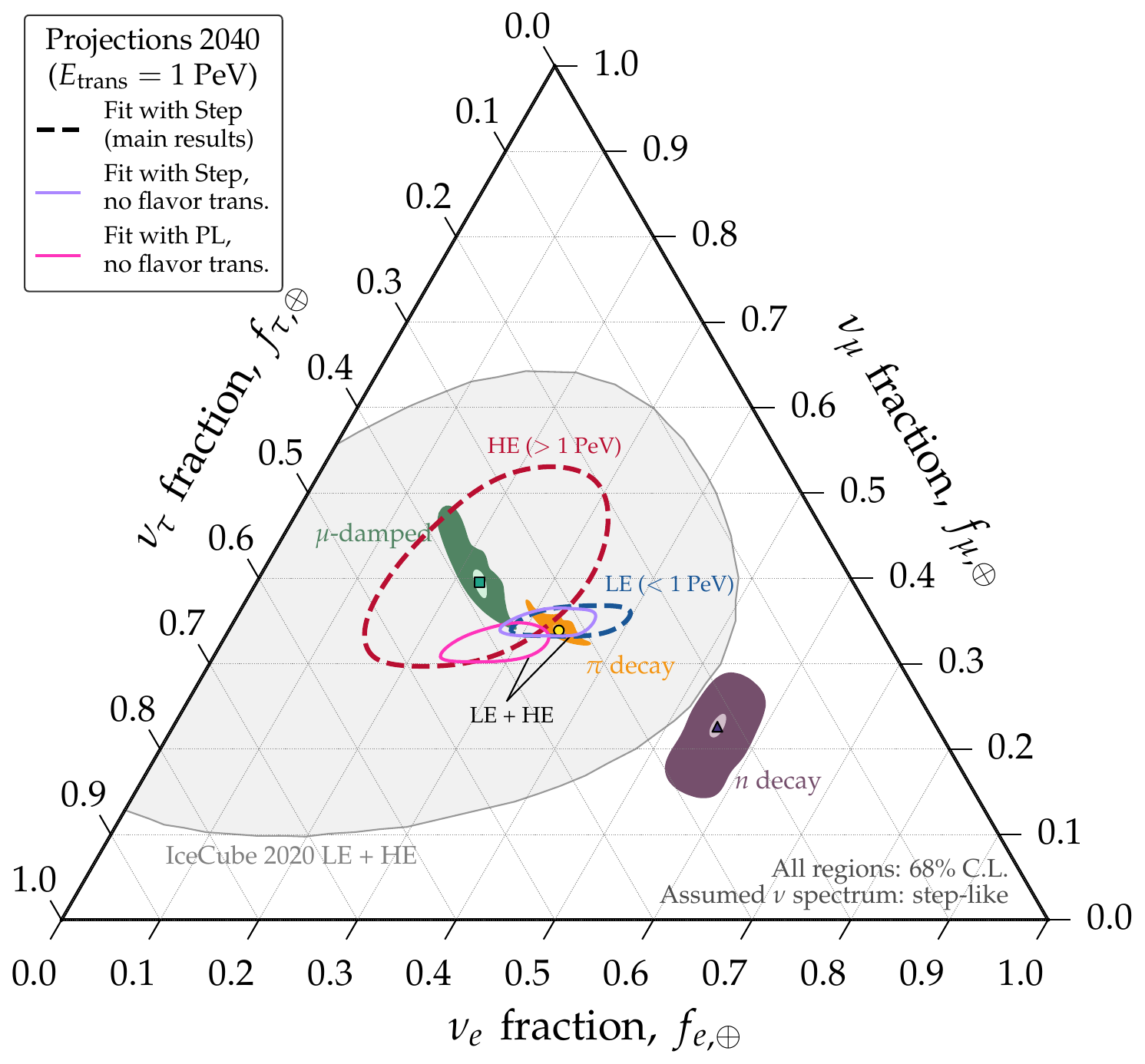}
 \caption{\textbf{\textit{Measuring the energy-averaged flavor composition when it contains a flavor transition in energy.}}  For this figure, projections are for the year 2040.  The neutrino spectrum is assumed to be step-like (Step), with a transition at 1~PeV (\figu{benchmark_flux}).  Our main results report the LE and HE flavor composition separately.  We contrast them against measurements of the energy-averaged flavor composition assuming step-like and single-power-law (PL) flux models without a flavor transition.  Previous energy-averaged measurements from IceCube and allowed flavor regions are the same as in \figu{triangle_hese_vs_tgm}.  See \Cref{sec:results_model-comparison} for details.  \textit{Flavor measurements must use flexible descriptions of the neutrino spectrum to avoid reporting inaccurate flavor-composition measurements.}}
\label{fig:compare_avg_measurement}
\end{figure}

Figure~\ref{fig:compare_avg_measurement} illustrates the measurement of the energy-averaged flavor composition in a situation when in reality it features a transition in energy.  In \figu{compare_avg_measurement}, the true flux is taken to be a Step spectrum with a transition at 1~PeV.  Our main results, duplicated in \figu{compare_avg_measurement} from \figu{triangles}, fit the observed event sample assuming also a Step spectrum, and measure the LE and HE flavor composition separately. Figure~\ref{fig:compare_avg_measurement} shows that the measurement of the energy-averaged flavor composition using the same event sample is affected by using a description of the neutrino flux that is unable to reproduce the true flux.  

First, fitting the observations using a Step spectrum but without allowing a flavor composition, \ie, fixing $f_{\alpha, \oplus}^{\rm LE} = f_{\alpha, \oplus}^{\rm HE}$, improves the measurement precision of the energy-averaged flavor composition compared to our main results for LE and HE, but erases the evidence for the muon-damped flavor composition at high energy.  More extremely, fitting the observations with a PL spectrum without allowing a flavor composition yields a preferred energy-averaged flavor composition that matches neither the pion-decay nor the muon-damped benchmarks.  Thus, \figu{compare_avg_measurement} points out the need to use a flexible description of the neutrino spectrum when measuring the flavor composition, even the energy-averaged one.


\subsection{Flavor composition at the sources}
\label{sec:results_flavor-sources}

Finally, we turn to the reconstruction of the flavor composition at the sources, using the method from \Refe~\cite{Bustamante:2019sdb} (see also \Refe~\cite{Song:2020nfh}).  As stated in \Cref{sec:analysis_flavor-sources}, we assume  there is no $\nu_\tau$ production in the sources, and so we infer only the LE and HE fractions of $\nu_e$ at the sources, $f_{e, {\rm S}}^{\rm LE}$ and $f_{e, {\rm S}}^{\rm HE}$, which completely determine the flavor composition at the sources, \ie, $\left(f_{e, {\rm S}}^{\rm LE}, f_{\mu, {\rm S}}^{\rm LE} \equiv 1 - f_{e, {\rm S}}^{\rm LE}, f_{\tau, {\rm S}}^{\rm LE} \equiv 0 \right)$, and similarly for HE. 

\begin{figure*}[th!]
 \centering
 \subfigure{\includegraphics[width=1.0\textwidth]{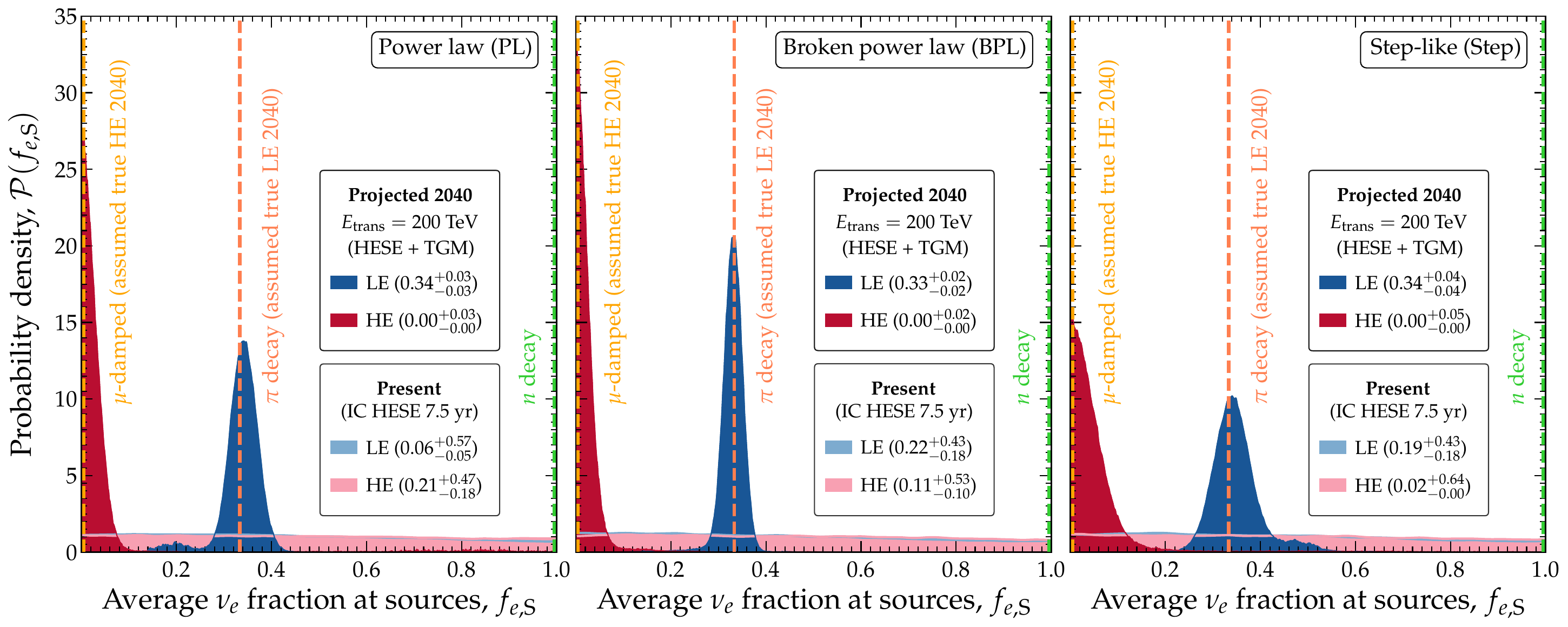}}
 \subfigure{\includegraphics[width=1.0\textwidth]{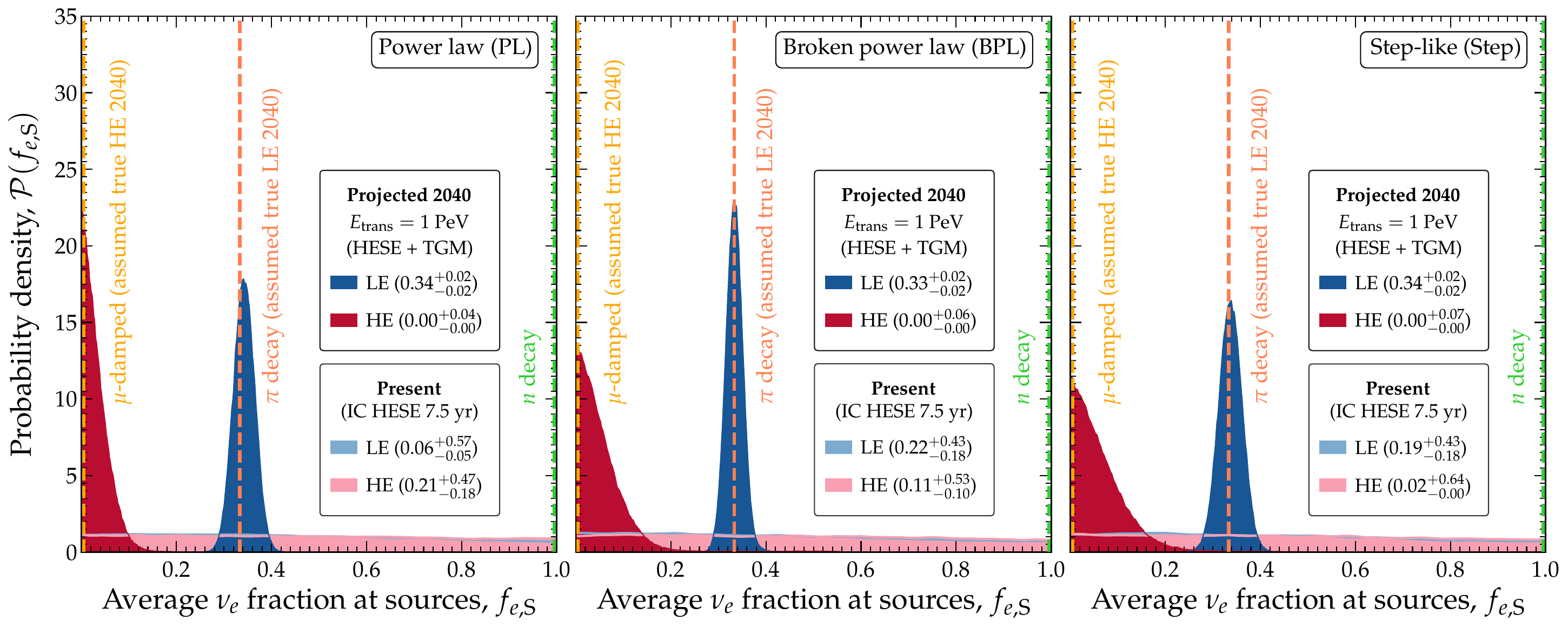}}\\
 \caption{\textbf{\textit{Inferred posterior distribution of the fraction of high-energy astrophysical neutrinos produced as $\nu_e$, at LE and HE}}.  Present results are from the IceCube 7.5-year HESE sample~\cite{IceCube:2020wum, IC75yrHESEPublicDataRelease}, using the presently allowed values of the neutrino mixing parameters from \texttt{NuFIT}~5.1~\cite{Esteban:2020cvm, nufit5.1}. Projected results are for the year 2040 and use the combined, cumulative detection of HESE and TGM events by multiple neutrino telescopes (Table~\ref{tab:detectors}), and projected measurements of the mixing parameters (Table~\ref{tab:parameters}).  The projections are made for three benchmark neutrino spectra (\figu{benchmark_flux})---PL (\textit{left}), BPL (\textit{center}), and Step (\textit{right})---which transition from pion decay at LE to muon-damped at HE, \ie, below and above $E_\textrm{trans}$.  We show results for $E_\textrm{trans} = 200$~TeV (\textit{top}) and 1~PeV (\textit{bottom}) .  The allowed intervals of flavor composition are at 68\%~C.L.  See Table~\ref{tab:results_sources} for further results and \Cref{sec:results_flavor-sources} for details.  \textit{There is no evidence of a flavor transition with energy in present HESE data, but projections reveal sensitivity to differentiate between the pion-decay, muon-damped, and neutron-decay neutrino production mechanisms at LE and HE.}}
 \label{fig:posterior_src}
\end{figure*}

Figure~\ref{fig:posterior_src} shows the resulting posteriors of $f_{e, {\rm S}}^{\rm LE}$ and $f_{e, {\rm S}}^{\rm HE}$.  Table~\ref{tab:results_sources} in Appendix~\ref{app:posteriors} shows the allowed intervals.  Our present-day results, generated assuming the 7.5-year IceCube HESE sample and the \texttt{NuFIT}~5.1~\cite{Esteban:2020cvm, nufit5.1} distributions of the neutrino mixing parameters (Table~\ref{tab:parameters}) are rather flat posteriors for LE and HE for all three benchmark flux models.  Therefore, presently, there is no strong preference for any specific flavor composition at LE or HE at the sources, nor for the existence of a flavor transition.  

This contrasts with the results of \Refes~\cite{Bustamante:2019sdb, Song:2020nfh}, which reported mild preference for the energy-averaged flavor composition at the sources being muon-damped, and a clear rejection of the composition from neutron decay.  In our analysis, similar conclusions cannot be reached at present for LE and HE separately, due primarily to splitting the limited number of HESE events among them.  

Figure~\ref{fig:posterior_src} also shows projections for the year 2040, generated assuming the combined detection of HESE and TGM by multiple detectors (Table~\ref{tab:detectors}) and forecasts of higher-precision knowledge of the mixing parameters (Table~\ref{tab:parameters}).  The widths of the posterior distributions are sensitive to the value of $E_\textrm{trans}$, which determines the number of events in the LE and HE regions, similarly to \Cref{sec:results_flavor-earth}.  We show results assuming a true value of $E_\textrm{trans}$ of 200~TeV and 1~PeV. 

In our projections, the flavor composition is measured accurately, with best-fit values of $f_{e, {\rm S}}^{\rm LE} \approx 0.34$ and $f_{e, {\rm S}}^{\rm HE} \approx 0$, matching the true values of our pion-decay and muon-damped benchmarks (\figu{benchmark_flux}), and a 68\%~C.L.~precision of 6--12\%, depending on the flux model.  Figure~\ref{fig:posterior_src} shows that this level of precision is sufficient to distinguish between the flavor composition at LE and HE, and to firmly disfavor the flavor composition coming from neutron decay, \ie, $\left( 1:0:0 \right)_{\rm S}$, in both energy regions.  

Like for the flavor composition at Earth (\Cref{tab:results_earth}), the measurement uncertainty of the flavor composition at the sources is comparable for the PL, BPL, and Step flux models, but the precision on $E_\textrm{trans}$ is different in each of them, though better overall.  At 68\%~C.L., it is 12\% for PL, 2\% for BPL, and 2\% for Step (Table~\ref{tab:results_sources}).  Compared to the measurement of the flavor composition at the Earth, the precision of the inferred flavor composition at the sources is better as a result of having to fit a single flavor fraction at LE and HE, rather than two.

\section{Envisioned improvements}
\label{sec:improvements}

Our methods and results are sound and informed by realistic detector capabilities, but we envision opportunities for improvement.  Below, we list three salient ones.

\begin{description}
 \item[Refined treatment of future detectors]
  Our analysis above relies on current IceCube event-reconstruction capabilities, and assume that future detectors will have different sizes but detection capabilities comparable to IceCube.  However, in reality, the capabilities of a future detector depend on its components, layout, and detection medium, which we do not account for. For example, in IceCube-Gen2, a 250-m spacing between detector strings would make HESE detection efficient above 200~TeV, rather than above 60~TeV as in IceCube, which has a 125-m spacing~\cite{IceCube-Gen2:2020qha}.  Further, the timeline we have adopted for upcoming detectors is tentative and subject to change, and may be augmented by even larger detectors, like the recently proposed HUNT~\cite{Huang:2023mzt}.  It may unfurl at a different---and, we hope, faster---rate.  
  Our projections may be refined as updated information becomes publicly available on the design, capabilities, and timeline of future detectors. 
 \item[Using lower-energy events]
  In our analysis, we have used HESE and TGM events, which have energies larger than 60~TeV and 100~GeV, respectively.  Including events with lower energies would not only increase the combined size of the event sample, but extend the energy range of the analysis down to 1~TeV.  For instance, preliminary studies show that using IceCube Medium-Energy Starting Events (MESE) would yield more precise flavor measurements than HESE at LE~\cite{IceCube:2023htx}. 
  However, since there is presently no MC event sample of lower-energy events publicly available, we do not include them in our analysis for now. 
 \item[Progress in event reconstruction]
  Measurements of the flavor composition may be improved by ongoing progress in event reconstruction~\cite{IceCube:2021umt, IceCube:2022njh, IceCube:2023ame} and advances in flavor identification based on using dedicated templates~\cite{IceCube:2020fpi, IceCube:2023fgt} and muon and neutron echoes~\cite{Li:2016kra, Steuer:2017tca, Farrag:2023jut}.
\end{description}


\section{Summary and outlook}
\label{sec:conclusion}

The flavor composition of TeV--PeV astrophysical neutrinos, \ie, the relative proportion of $\nu_e$, $\nu_\mu$, and $\nu_\tau$ in their flux, plays an important role in revealing the neutrino production mechanisms, the features and identities of their largely unknown astrophysical sources, and potential new physics.  Due to the difficulty in identifying flavor in high-energy neutrino telescopes and to a limited number of detected neutrinos, previous studies have measured the flavor composition at Earth under the assumption that it is constant over the observed range of neutrino energies, or averaging it over this range. 

Yet, in actuality, the flavor composition likely varies with neutrino energy, as different physical processes driving neutrino production, with different yields of $\nu_e$, $\nu_\mu$, and $\nu_\tau$, act at different energies, or as new-physics effects become relevant at different energies.  When measuring the energy-averaged flavor composition, the above effects are washed out and are harder to spot, eroding the insight on them that we can hope to extract.

To remedy this, we have reported the first measurements of the energy dependence of the high-energy neutrino flavor composition.  Our methods build on the experience of previous flavor measurements, and rely on nuanced modeling of the detection capabilities of IceCube to provide robust  results based on present-day data and informed projections for the next two decades.  Today, the measurement is challenging, due to the  limited number of detected events.  In the near future, the growing number of events detected by upcoming neutrino telescopes is bound to increase the precision substantially.  

We have assessed the capability to measure a transition in the flavor composition from low to high neutrino energy.  Rather than studying specific models of neutrino production or new physics, we have explored three benchmark models of the neutrino energy spectrum that represent the breadth of predictions: a single unbroken power law (PL), a broken power law (BPL), and a step-like spectrum (Step) with a large change in flux normalization.  Each model includes a transition in the flavor composition from low to high energy.  In our projections, we assume the flavor composition at LE to be given by the full pion decay chain, $\left( 1:2:0 \right)_{\rm S}$, and, at high energies, by muon-damped pion decay, $\left( 0:1:0 \right)_{\rm S}$.  In the BPL and Step models, the flavor transition happens concomitantly with a break in the power-law spectrum or a change in its normalization, respectively.

For our present-day results, we use the publicly available IceCube 7.5-year sample of High-Energy Starting Events (HESE).  Because of the paucity of events in the sample, we are unable to identify the presence of a flavor transition.  Unsurprisingly, the error in the flavor composition measured at low and high energy covers nearly the full space of possibilities, even at 68\%~C.L.

For our projections, we use the detection of HESE events plus through-going muons (TGM) by IceCube and by future neutrino telescopes Baikal-GVD, IceCube-Gen2, KM3NeT, P-ONE, TAMBO, and TRIDENT.  Using multiple telescopes, most of which are larger than IceCube, boosts the detection rate of HESE.  Adding TGM events improves the measurement of the $\nu_\mu$ content.  

By the year 2040, it might be possible to identify a flavor transition at an energy of 1~PeV and, marginally, to distinguish between the flavor composition at low and high energies, thanks to a reduction in the uncertainty of its measurement in a factor of 5--6.  The distinction is marginal due to the rarity of events above 1~PeV.  The significance of these claims improves if the transition occurs instead at 200~TeV, which entails a more even splitting of events between low and high energies.  However,  identifying a transition in the flavor composition without relying on identifying also an associated change in the spectrum shape might require observations beyond 2040.

The above measurements of the flavor composition at Earth, combined with projected high-precision knowledge of the neutrino mixing parameters, would allow us to infer the flavor composition with which neutrinos are produced at their sources.  By 2040, we could measure the low- and high-energy flavor composition accurately, with a precision of 6--12\%.  This is enough to establish the full pion decay as the main neutrino production mechanism at LE and muon-damped production at HE, and to discard alternatives like production via neutron decay.

Overall, our findings show that the measurement of the flavor composition at different energies is challenging today, but is set to improve appreciably in the near future.  Pursuing it will strengthen and broaden the insight into neutrino astrophysics and fundamental physics that flavor measurements deliver.

\acknowledgements

The authors would like to thank Paschal Coyle, Francis Halzen, Elisa Resconi, and V\'eronique van Elewyck, and Donglian Xu for helpful discussion. The authors also thank Fan Hu for providing TRIDENT effective areas. CAA is supported by the Faculty of Arts and Sciences of Harvard University and was partially supported by the Alfred P. Sloan Foundation in this work, and by the NSF CAREER Award PHY-2239795. MB and DFGF are supported by the Villum Fonden under Project No.\ 29388. This project has received funding from the European Union’s Horizon 2020 research and innovation program under the Marie Sklodowska-Curie Grant Agreement No.~847523 ‘INTERACTIONS’. QL and ACV are supported by the Arthur B. McDonald Canadian Astroparticle Physics Research Institute, with equipment funded by the Canada Foundation for Innovation and the Province of Ontario, and housed at the Queen’s Centre for Advanced Computing. Research at Perimeter Institute is supported by the Government of Canada through the Department of Innovation, Science, and Economic Development, and by the Province of Ontario. ACV acknowledges further support from NSERC and the Ontario Ministry of Colleges and Universities. NS is supported by the National Natural Science Foundation of China (NSFC) Project No. 12047503. NS also acknowledges the UK Science and Technology Facilities Council for support through the Quantum Sensors for the Hidden Sector collaboration under the grant ST/T006145/1. 


\bibliography{references}

\appendix


\section{Posterior distributions and allowed intervals of the model parameters}
\label{app:posteriors}

\renewcommand{\theequation}{A\arabic{equation}}
\renewcommand{\thefigure}{A\arabic{figure}}
\renewcommand{\thetable}{A\arabic{table}}
\setcounter{figure}{0} 
\setcounter{table}{0} 

In Section~\ref{sec:analysis} in the main text, we describe how we compute the joint posterior probability of the free model parameters (Table~\ref{tab:parameters}) that we use to measure the flavor composition at Earth, $f_{\alpha, \oplus}^{\rm LE}$ and $f_{\alpha, \oplus}^{\rm HE}$, and infer the flavor composition at the sources, $f_{\alpha, {\rm S}}^{\rm LE}$ and $f_{\alpha, {\rm S}}^{\rm HE}$.  In the main text, we show posteriors of the flavor composition, marginalized over all the other model parameters (Figs.~\ref{fig:triangle_main}, \ref{fig:triangles}, and \ref{fig:posterior_src}). 

To complement those results, and for the purpose of illustration, below we show joint two-dimensional pair-wise posteriors for all the model parameters for our present and projected results for the year 2040.  For our projections, we fix the transition energy to $E_\textrm{trans} = 1$~PeV.  (We have also generated projections assuming $E_\textrm{trans} = 200$~TeV, and for other years, but we do not show the joint posteriors for those cases.)

\begin{figure*}
 \centering
 \includegraphics[width=1.0\textwidth]{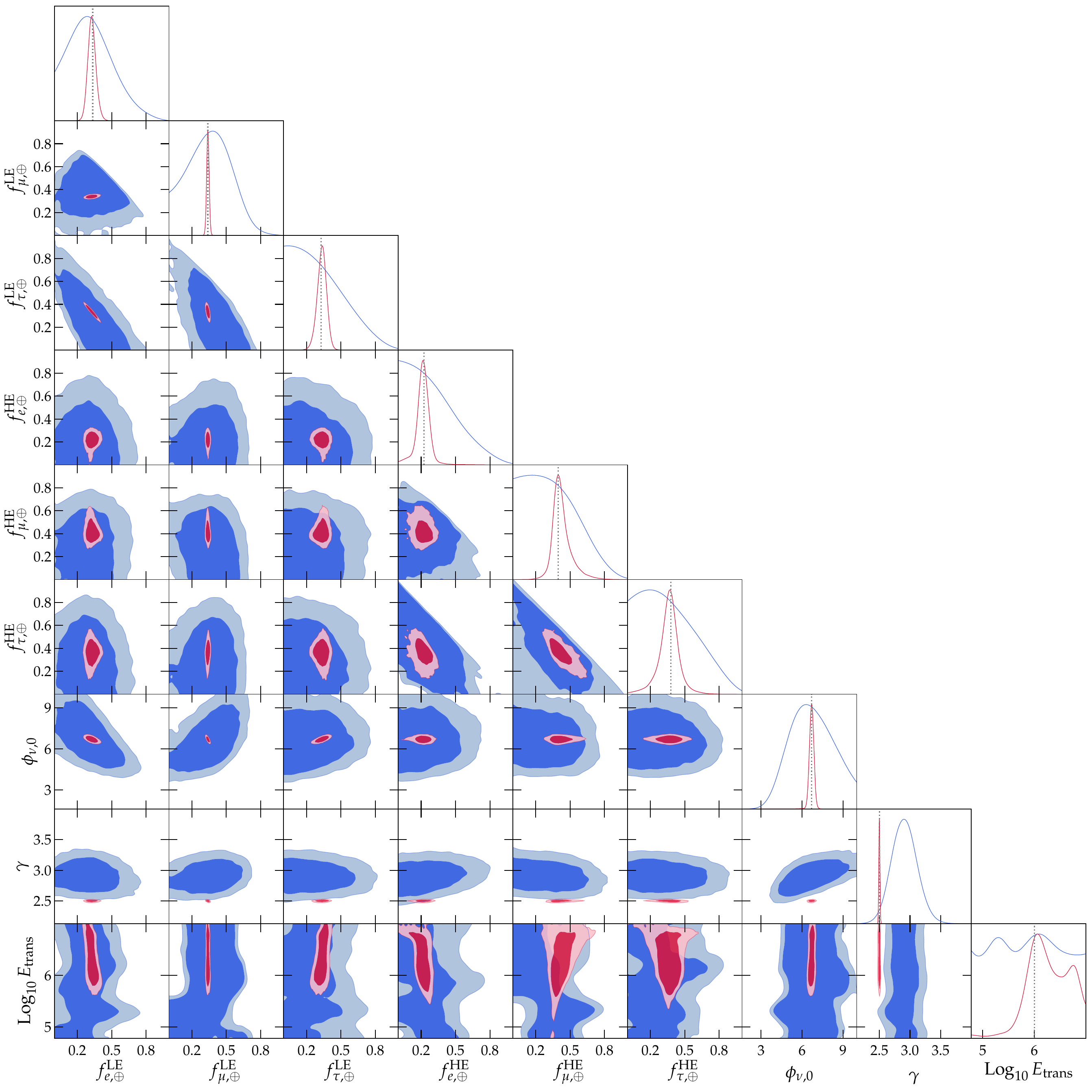}\\
 \includegraphics[width=0.7\textwidth]{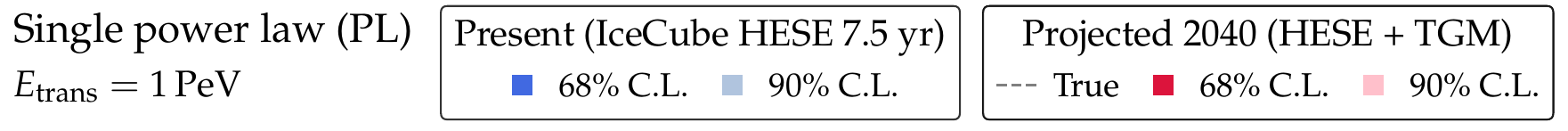}
 \caption{\textbf{\textit{Joint posterior distributions of the model parameters in the measurement of the flavor composition at Earth, for the power law (PL) benchmark neutrino spectrum.}}  See Table~\ref{tab:parameters} for a description of the parameters, including units, and \figu{benchmark_flux} for the specific PL flux that we use to make our projections.  The full joint posterior is \equ{posterior} in the main text.  Each panel shows the two-dimensional posterior marginalized over all model parameters except for the two in the panel.  Allowed regions are for 68\% and 90\% C.L..  In this figure (and also in Figs.~\ref{fig:bpl_corner} and \ref{fig:step_corner}), we fix the true value of the transition energy in our projections to $E_\textrm{trans} = 1$~PeV.   See Table~\ref{tab:results_earth} for best-fit values and one-dimensional allowed intervals of selected parameters.  \Cref{sec:results_flavor-earth} in the main text for details.}
 \label{fig:pl_corner}
\end{figure*}

\begin{figure*}
 \centering
 \includegraphics[width=1.0\textwidth]{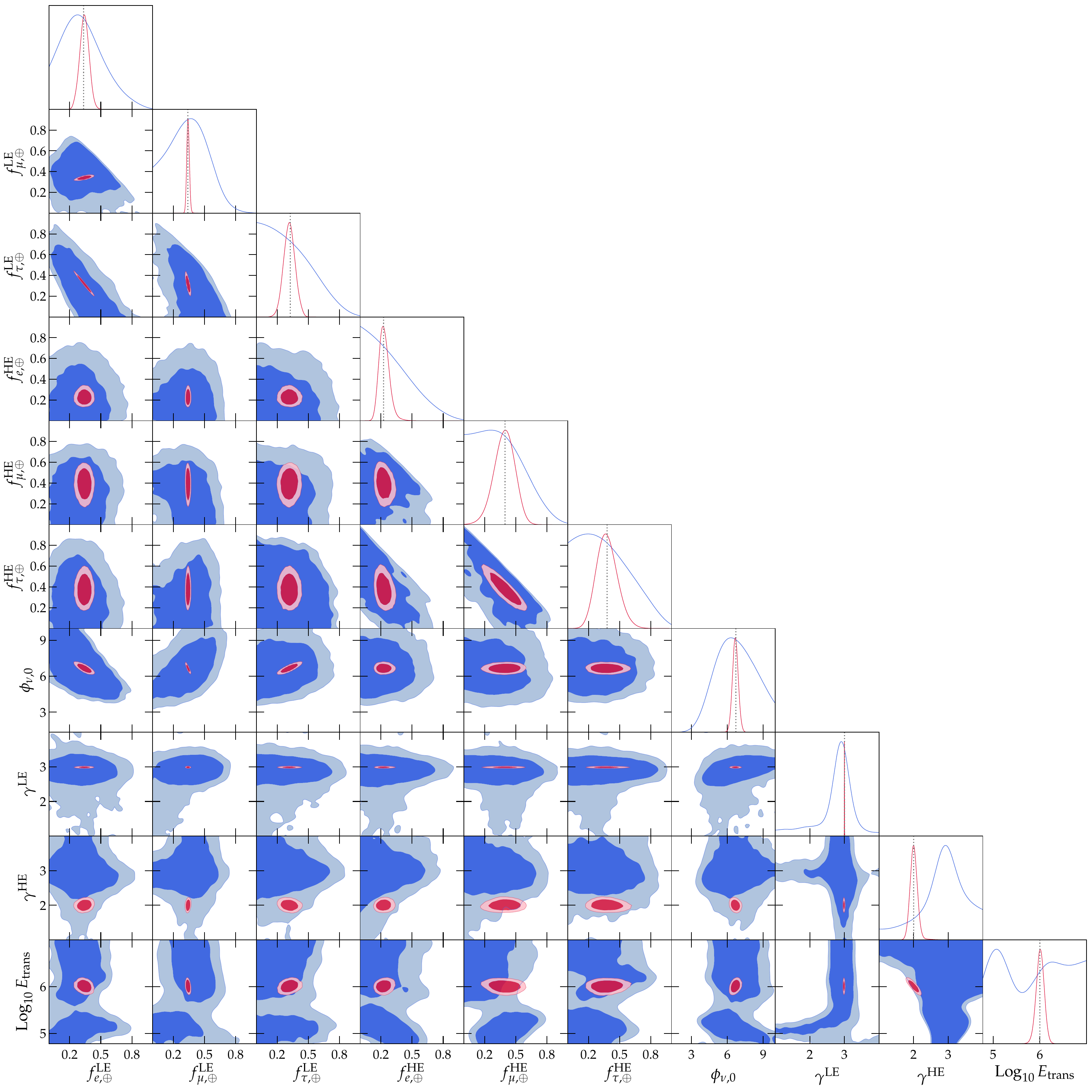}\\
 \includegraphics[width=0.7\textwidth]{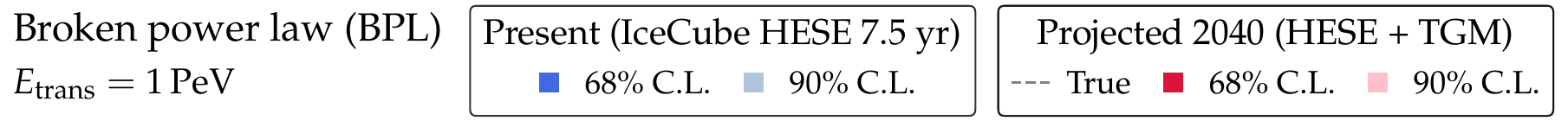}
 \caption{\textbf{\textit{Joint posterior distributions of the model parameters in the measurement of the flavor composition at Earth, for the broken power law (BPL) benchmark neutrino spectrum.}}  Same as \figu{pl_corner}, but for the BPL flux.}
 \label{fig:bpl_corner}
\end{figure*}

\begin{figure*}
 \centering
 \includegraphics[width=1.0\textwidth]{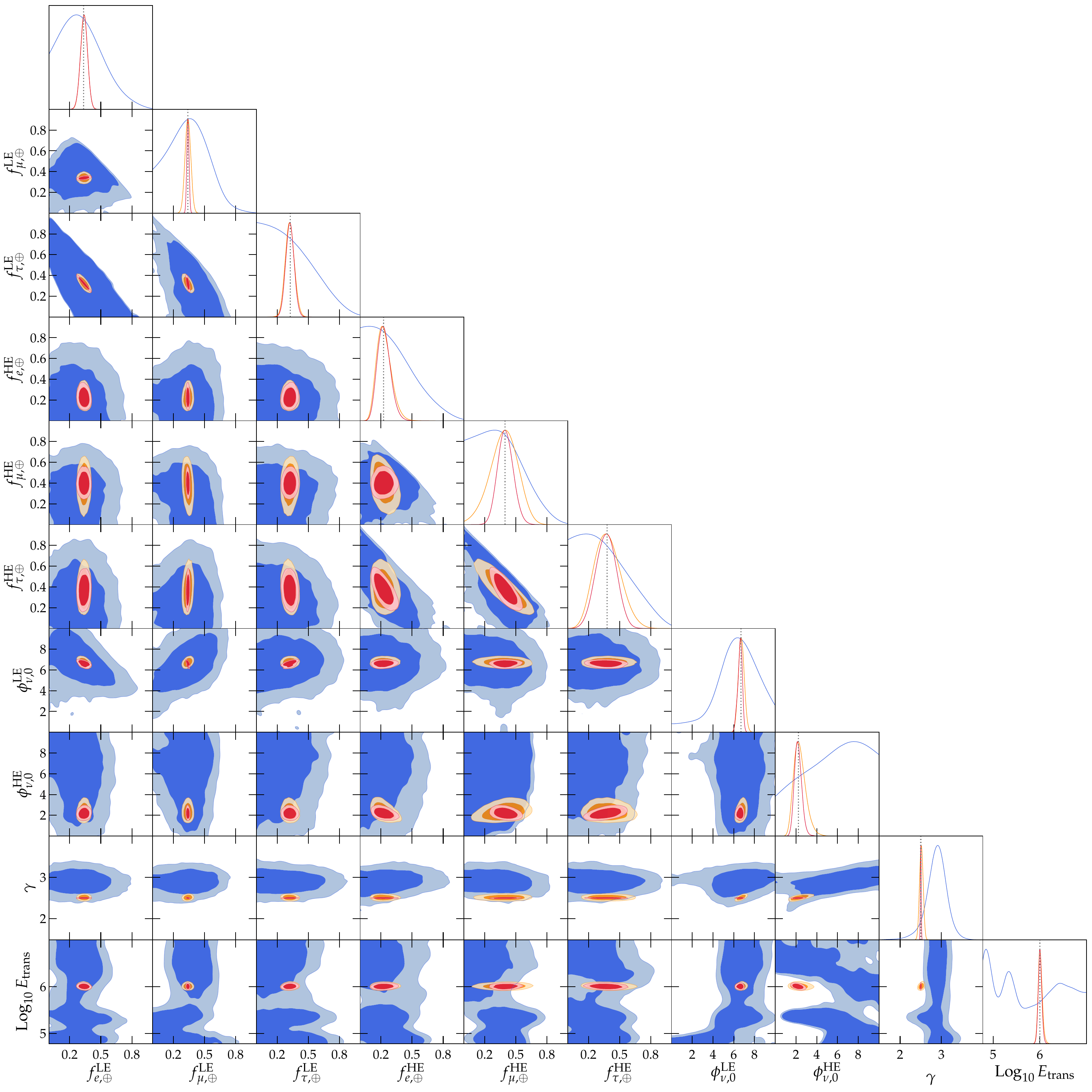}\\
 \includegraphics[width=0.7\textwidth]{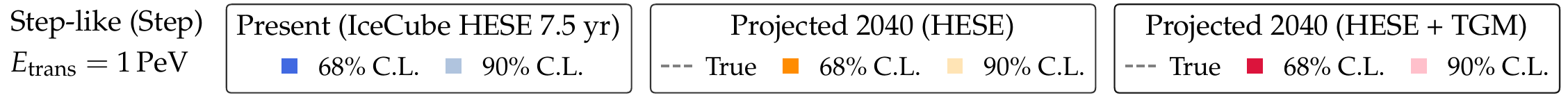}
 \caption{\textbf{\textit{Joint posterior distributions of the model parameters in the measurement of the flavor composition at Earth, for the step-like (Step) benchmark neutrino spectrum.}}  Same as \figu{pl_corner}, but for the Step flux. Furthermore, the projection for 2040 with HESE-only selection is shown as orange regions, where the improvement from the inclusion of TGM can be seen compared to the red regions.}
 \label{fig:step_corner}
\end{figure*}

Figures~\ref{fig:pl_corner}--\ref{fig:step_corner} show posteriors for the measurement of the flavor composition at Earth for our PL, BPL, and Step benchmark flux models.

\begin{figure*}
 \centering
 \includegraphics[width=1.0\textwidth]{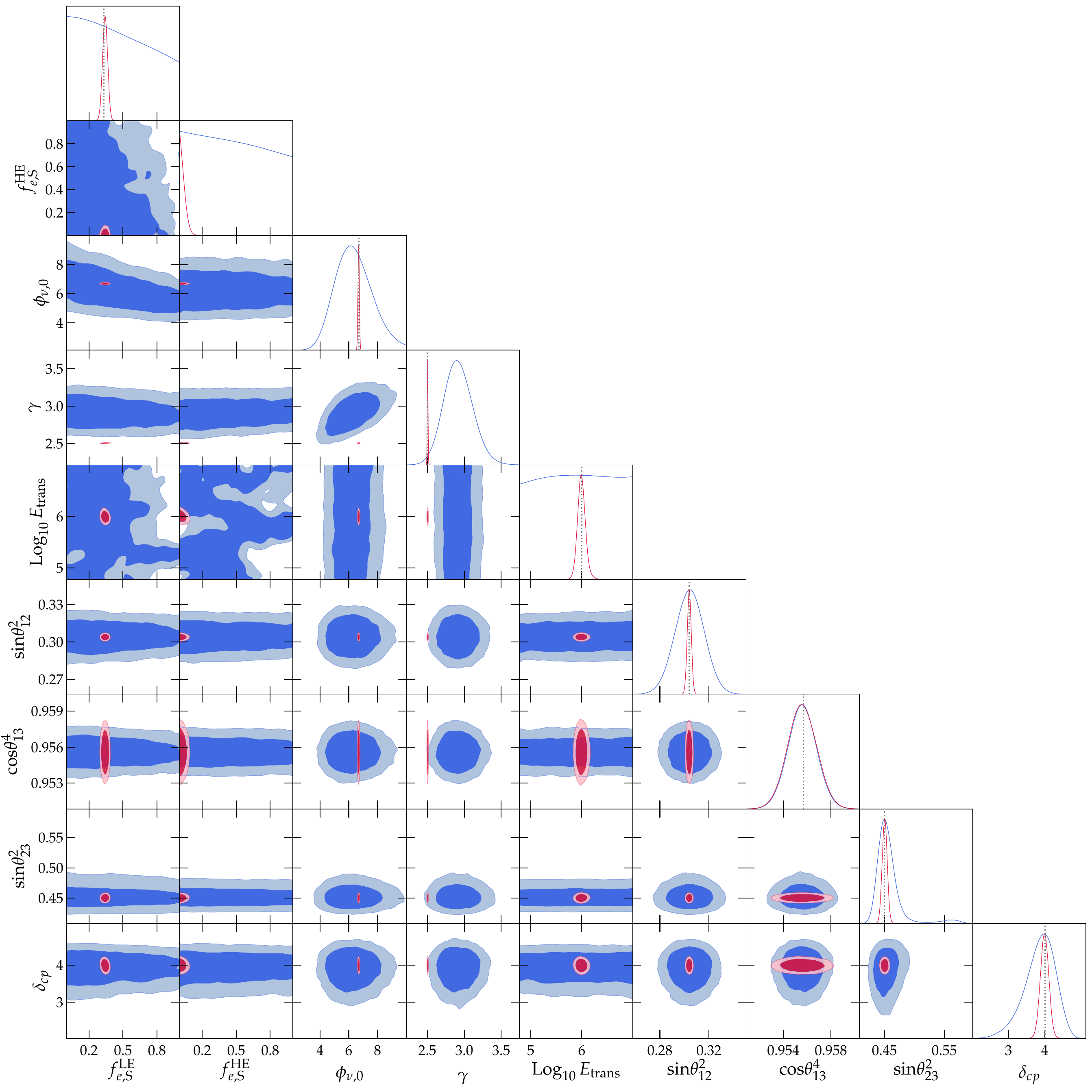}\\
 \includegraphics[width=0.7\textwidth]{pl_corner_label.pdf}
 \caption{\textbf{\textit{Joint posterior distributions of the model parameters inferring the flavor composition at sources, for the power law (PL) benchmark neutrino spectrum.}}  See Table~\ref{tab:parameters} for a description of the parameters, including units, and \figu{benchmark_flux} for the specific PL flux that we use to make our projections.  The full joint posterior is \equ{posterior} in the main text.  Each panel shows the two-dimensional posterior marginalized over all model parameters except for the two in the panel.  Allowed regions are for 68\% and 90\% C.L..  In this figure (and also in Figs.~\ref{fig:bpl_corner_src} and \ref{fig:step_corner_src}), we fix the true value of the transition energy in our projections to $E_\textrm{trans} = 1$~PeV.  See Table~\ref{tab:results_sources} for best-fit values and one-dimensional allowed intervals of selected parameters.  See \Cref{sec:results_flavor-sources} in the main text for details.}
 \label{fig:pl_corner_src}
\end{figure*}

\begin{figure*}
 \centering
 \includegraphics[width=1.0\textwidth]{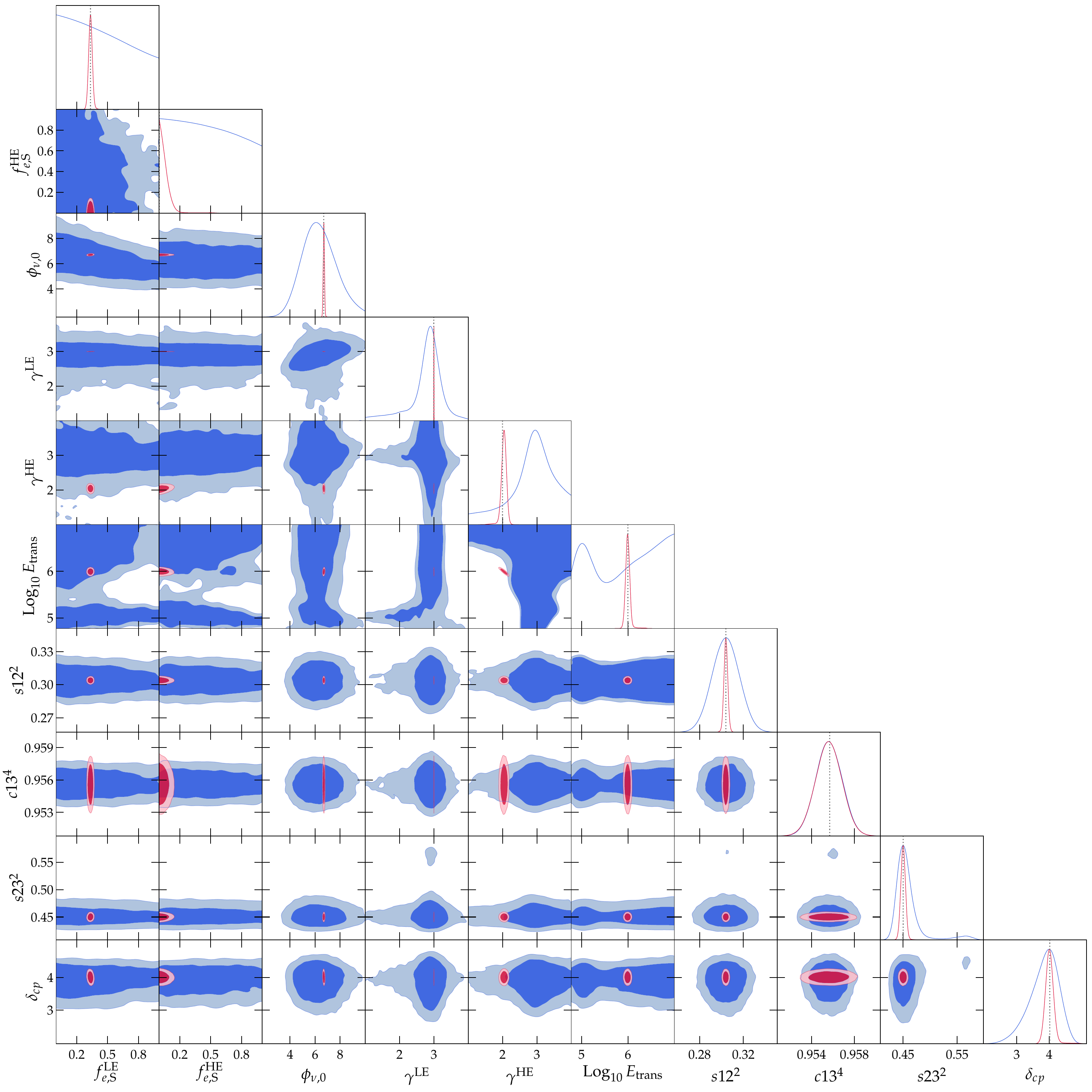}\\
 \includegraphics[width=0.7\textwidth]{bpl_corner_label.pdf}
 \caption{\textbf{\textit{Joint posterior distributions of the model parameters inferring the flavor composition at sources, for the broken power law (BPL) benchmark neutrino spectrum.}}  Same as \figu{pl_corner_src}, but for the BPL flux.}
 \label{fig:bpl_corner_src}
\end{figure*}

\begin{figure*}
 \centering
 \includegraphics[width=1.0\textwidth]{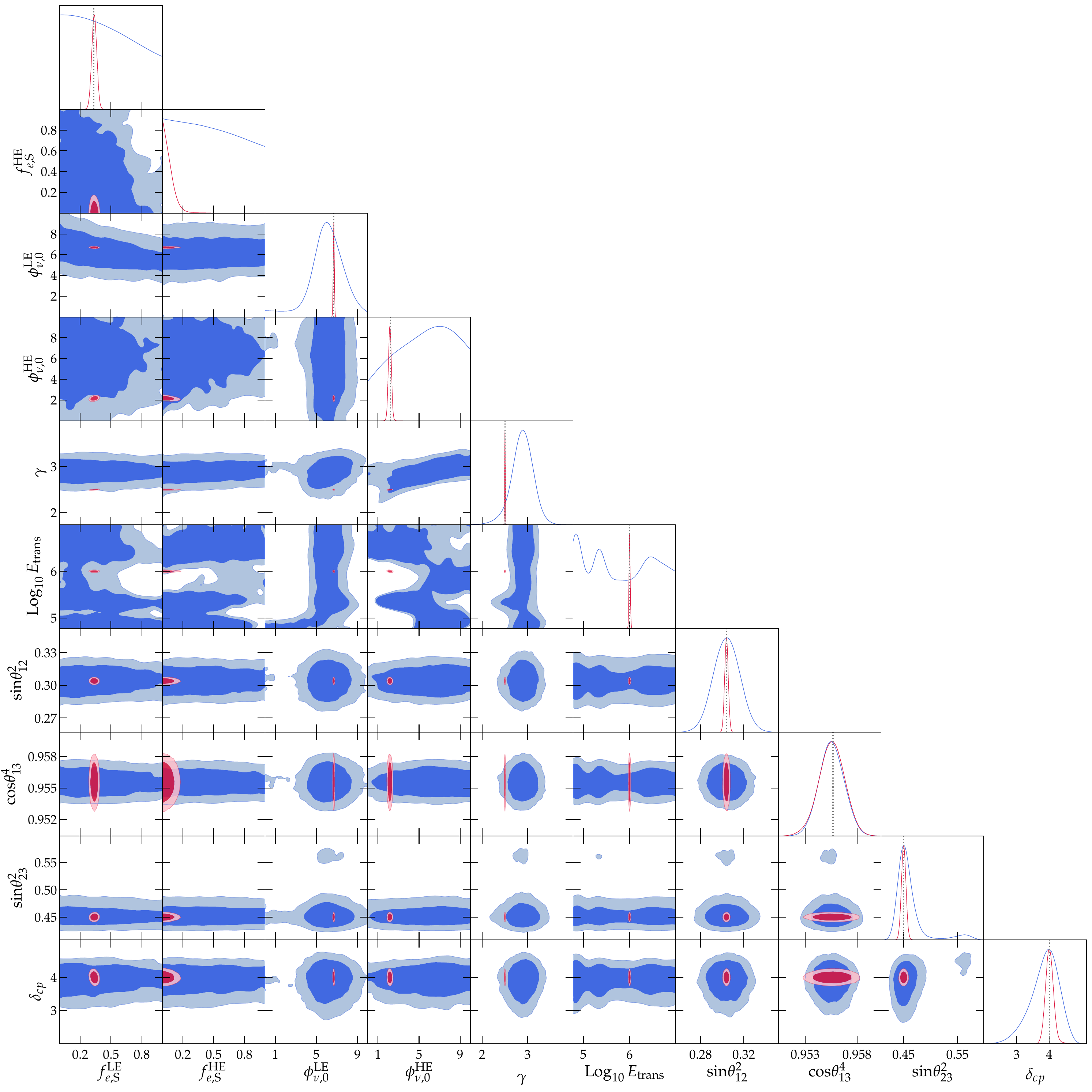}\\
 \includegraphics[width=0.7\textwidth]{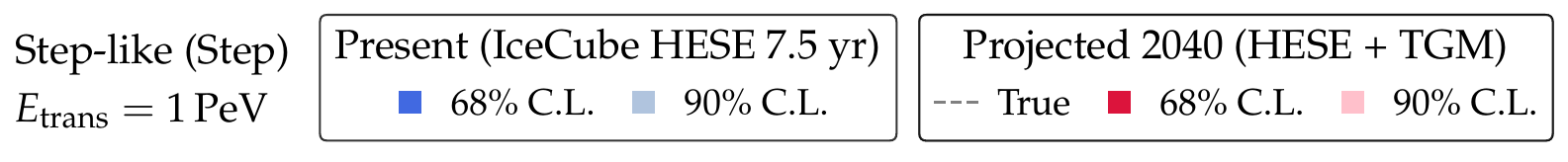}
 \caption{\textbf{\textit{Joint posterior distributions of the model parameters inferring the flavor composition at sources, for the step-like (Step) benchmark neutrino spectrum.}}  Same as \figu{pl_corner_src}, but for the Step flux.}
 \label{fig:step_corner_src}
\end{figure*}

Figures~\ref{fig:pl_corner_src}--\ref{fig:step_corner_src} show posteriors for the inferred flavor composition at the sources for our PL, BPL, and Step benchmark flux models.

Figures \ref{fig:pl_corner}--\ref{fig:step_corner_src} confirm that there are appreciable correlations between the flavor fractions; this is shown already in Figs.~\ref{fig:triangle_main}, \ref{fig:triangles}, and \ref{fig:posterior_src} in the main text.  The figures also reveal that the parameters that determine the shape of the neutrino spectrum (Table~\ref{tab:parameters})---the flux normalization factors, spectral indices, and transition energy---are only weakly correlated with the flavor fractions, especially in our projections.  This lends support to our finding  (\Cref{sec:results}) that the evidence for the existence of a transition in energy stems largely from a change in the spectrum shape.

Table~\ref{tab:results_earth} shows the best-fit values of the  flavor composition measured at Earth, $f_{\alpha, \oplus}^{\rm LE}$ and $f_{\alpha, \oplus}^{\rm HE}$, and the allowed ranges of the transition energy, for each of our benchmark flux models.  For the flavor composition, given the significant correlations between the flavor fractions, instead of showing one-dimensional marginalized results, we refer to their joint two-dimensional allowed regions in Figs.~\ref{fig:pl_corner}--\ref{fig:step_corner}, and Figs.~\ref{fig:triangle_main} and \ref{fig:triangles} in the main text.

Table~\ref{tab:results_sources} shows the allowed ranges of the inferred flavor composition at the sources, $f_{e, {\rm S}}^{\rm LE}$ and $f_{e, {\rm S}}^{\rm HE}$, and of the transition energy, for each of our benchmark flux models.  

\begingroup
\squeezetable
\begin{table*}[t!]
 \begin{ruledtabular}  
  \caption{\label{tab:results_earth}\textbf{\textit{Flavor composition of high-energy astrophysical neutrinos at Earth, measured from fits to present IceCube data and projected data.}}  Present measurements are from the public IceCube 7.5-year HESE sample~\cite{IceCube:2020wum, IC75yrHESEPublicDataRelease} (Section~\ref{sec:analysis_hese}).  Forecasts are from the combined, cumulative detection of HESE and TGM (Sections~\ref{sec:analysis_tgm} and \ref{sec:analysis_hese-tgm}) in the year 2040 by IceCube and future neutrino telescopes Baikal-GVD, IceCube-Gen2, KM3NeT, P-ONE, TAMBO, and TRIDENT (Section~\ref{sec:detectors}).  The flavor composition is measured separately for our three benchmark neutrino spectra (Section~\ref{sec:scenarios_production})---power law (PL), broken power law (BPL), and step-like (Step)---using the methods in Section~\ref{sec:analysis_flavor-earth}, with the priors from Table~\ref{tab:parameters}.  We have converted the flavor angles that are varied in the fit (Table~\ref{tab:parameters}) into the flavor fractions at LE, $f_{\alpha, \oplus}^{\rm LE}$ ($\alpha = e, \mu, \tau)$, and high energies, $f_{\alpha, \oplus}^{\rm HE}$; the conversion is in Section~\ref{sec:analysis_flavor-earth}.  For each parameter, its allowed range is the one-dimensional 68\%~C.L. obtained by marginalizing the multi-dimensional posterior, \equ{posterior}, over all other parameters.  See Section~\ref{sec:results_flavor-earth} for details.}
  \centering
  \renewcommand{\arraystretch}{1.35}
  \begin{tabular}{cccccccccccc}
   \multirow{4}{*}{\makecell{Flux\\model}}                           & 
   \multicolumn{3}{c}{Present (IceCube, HESE 7.5 yr)}     &
   \multicolumn{4}{c}{Projected 2040 (multiple detectors, HESE + TGM)}  \\
   \cline{2-4}
   \cline{5-8}
   &
   &
   \multicolumn{2}{c}{Flavor composition at Earth (best fit)\footnote{\label{fnote1}We show only the best-fit values of the flavor composition because the flavor fractions are highly correlated.  We refer instead to their two-dimension allowed ranges in Figs.~\ref{fig:pl_corner}--\ref{fig:step_corner}, and Figs.~\ref{fig:triangle_main} and \ref{fig:triangles} in the main text.}}      &
   &
   &
   \multicolumn{2}{c}{Flavor composition at Earth (best fit)\footref{fnote1}}      \\
   \cline{3-4}
   \cline{7-8}
   &
   $E_\textrm{trans}$ [TeV] &
   \multicolumn{1}{c}{LE ($< E_\textrm{trans}$)}        &
   \multicolumn{1}{c}{HE ($> E_\textrm{trans}$)}        &
   \multicolumn{2}{c}{$E_\textrm{trans}$ [TeV]}         &
   \multicolumn{1}{c}{LE ($< E_\textrm{trans}$)}        &
   \multicolumn{1}{c}{HE ($> E_\textrm{trans}$)}        \\
   \cline{5-6}
   &
   &
   $\left( f_{e, \oplus}^{\rm LE}, 
   f_{\mu, \oplus}^{\rm LE}, 
   f_{\tau, \oplus}^{\rm LE} \right)$                   &
   $\left( f_{e, \oplus}^{\rm HE}, 
   f_{\mu, \oplus}^{\rm HE}, 
   f_{\tau, \oplus}^{\rm HE} \right)$                   &
   True                                                 &
   Measured                                             &
   $\left( f_{e, \oplus}^{\rm LE}, 
   f_{\mu, \oplus}^{\rm LE}, 
   f_{\tau, \oplus}^{\rm LE} \right)$                   &
   $\left( f_{e, \oplus}^{\rm HE}, 
   f_{\mu, \oplus}^{\rm HE}, 
   f_{\tau, \oplus}^{\rm HE} \right)$                   \\
   \hline
   PL                                                   &
   $192.63_{-87.41}^{+2776.91}$                         &
   $\left( 0.28, 
   0.35, 
   0.37 \right)$                                 &
   $\left( 0.24, 
   0.32, 
   0.44 \right)$                                 &
   200                                                  &
   $197.69_{-51.06}^{+104.54}$                          &
   $\left( 0.31, 
   0.33, 
   0.36 \right)$                                 &
   $\left( 0.22, 
   0.40, 
   0.38 \right)$                                 \\
   &
   &
   &
   &
   $10^3$                                               &
   $893.32_{-396.29}^{+3068.38}$                        &
   $\left( 0.32, 
   0.34, 
   0.34 \right)$                                 &
   $\left( 0.21, 
   0.41, 
   0.38 \right)$                                 \\
   BPL                                                  &
   $115.83_{-48.78}^{+2205.10}$                         &
   $\left( 0.22, 
   0.29, 
   0.49 \right)$                                 &
   $\left( 0.12, 
   0.27, 
   0.61 \right)$                                 &
   200                                                  &
   $190.95_{-12.47}^{+16.50}$                           &
   $\left( 0.35, 
   0.33, 
   0.32 \right)$                                 &
   $\left( 0.22, 
   0.40, 
   0.38 \right)$                                 \\
   &
   &
   &
   &
   $10^3$                                               &
   $955.53_{-132.49}^{+255.22}$                         &
   $\left( 0.34, 
   0.32, 
   0.34 \right)$                                 &
   $\left( 0.21, 
   0.39, 
   0.40 \right)$                                 \\
   Step                                                 &
   $78.85_{-15.83}^{+2066.11}$                         &
   $\left( 0.24, 
   0.34, 
   0.42 \right)$                                 &
   $\left( 0.25, 
   0.32, 
   0.43 \right)$                                 &
   200                                                  &
   $198.57_{-9.91}^{+9.24}$                           &
   $\left( 0.33, 
   0.31, 
   0.36 \right)$                                 &
   $\left( 0.21, 
   0.39, 
   0.40 \right)$                                 \\
   &
   &
   &
   &
   $10^3$                                               &
   $985.33_{-62.98}^{+124.21}$                          &
   $\left( 0.34, 
   0.34, 
   0.32 \right)$                                 &
   $\left( 0.24, 
   0.41, 
   0.35 \right)$                                 \\
  \end{tabular}
 \end{ruledtabular}  
\end{table*}
\endgroup

\begingroup
\squeezetable
\begin{table*}[t!]
 \begin{ruledtabular}  
  \caption{\label{tab:results_sources}\textbf{\textit{Flavor composition of high-energy astrophysical neutrinos at the sources, inferred from fits to present IceCube data and projected data.}}  We assume no $\nu_\tau$ production in the sources, \ie, $f_{\tau, {\rm S}}^{\rm LE} = f_{\tau, {\rm S}}^{\rm HE} = 0$, so we only need to infer the $\nu_e$ fractions, $f_{e, {\rm S}}^{\rm LE}$ and $f_{e, {\rm S}}^{\rm HE}$; the $\nu_\mu$ fractions are $f_{\mu, {\rm S}}^{\rm LE} = 1 - f_{e, {\rm S}}^{\rm LE}$ and $f_{\mu, {\rm S}}^{\rm HE} = 1- f_{e, {\rm S}}^{\rm HE}$.  Present measurements are from the public IceCube 7.5-year HESE sample~\cite{IceCube:2020wum, IC75yrHESEPublicDataRelease} (Section~\ref{sec:analysis_hese}) and use present uncertainties on neutrino mixing parameters from \texttt{NuFIT}~5.1~\cite{Esteban:2020cvm, nufit5.1}, (Table~\ref{tab:parameters}).  Forecasts are from the combined, cumulative detection of HESE and TGM (Sections~\ref{sec:analysis_tgm} and \ref{sec:analysis_hese-tgm}) in the year 2040 by IceCube and future neutrino telescopes Baikal-GVD, IceCube-Gen2, KM3NeT, P-ONE, TAMBO, and TRIDENT (Section~\ref{sec:detectors}), and assume improved measurements of the mixing parameters by DUNE, Hyper-Kamiokande, and JUNO (Table~\ref{tab:parameters}).  The flavor composition is measured separately for our three benchmark neutrino spectra (Section~\ref{sec:scenarios_production})---power law (PL), broken power law (BPL), and step-like (Step)---using the methods in Section~\ref{sec:analysis_flavor-earth}, with the priors from Table~\ref{tab:parameters}.  For each parameter, its allowed range is the one-dimensional 68\%~C.L. obtained by marginalizing the multi-dimensional posterior, \equ{posterior}, over all other parameters.  See Section~\ref{sec:results_flavor-sources} for details.}
  \centering
  \renewcommand{\arraystretch}{1.35}
  \begin{tabular}{cccccccccccc}
   \multirow{4}{*}{\makecell{Flux\\model}}                           & 
   \multicolumn{3}{c}{Present (IceCube, HESE 7.5 yr)}     &
   \multicolumn{4}{c}{Projected 2040 (multiple detectors, HESE + TGM)}  \\
   \cline{2-4}
   \cline{5-8}
   &
   &
   \multicolumn{2}{c}{Flavor composition at sources\footnote{\label{fnote1}Assuming $f_{\tau, {\rm S}}^{\rm LE} = f_{\tau, {\rm S}}^{\rm HE} = 0$, so $f_{\mu, {\rm S}}^{\rm LE} = 1 - f_{e, {\rm S}}^{\rm LE}$ and $f_{\mu, {\rm S}}^{\rm HE} = 1 - f_{e, {\rm S}}^{\rm HE}$.}}      &
   &
   &
   \multicolumn{2}{c}{Flavor composition at sources\footref{fnote1}}      \\
   \cline{3-4}
   \cline{7-8}
   &
   $E_\textrm{trans}$ [TeV] &
   \multicolumn{1}{c}{LE ($< E_\textrm{trans}$)}        &
   \multicolumn{1}{c}{HE ($> E_\textrm{trans}$)}        &
   \multicolumn{2}{c}{$E_\textrm{trans}$ [TeV]}         &
   \multicolumn{1}{c}{LE ($< E_\textrm{trans}$)}        &
   \multicolumn{1}{c}{HE ($> E_\textrm{trans}$)}        \\
   \cline{5-6}
   &
   &
   $\left( f_{e, {\rm S}}^{\rm LE}, 
   f_{\mu, {\rm S}}^{\rm LE}, 
   f_{\tau, {\rm S}}^{\rm LE} \right)$                  &
   $\left( f_{e, {\rm S}}^{\rm HE}, 
   f_{\mu, {\rm S}}^{\rm HE}, 
   f_{\tau, {\rm S}}^{\rm HE} \right)$                  &
   True                                                 &
   Measured                                             &
   $\left( f_{e, {\rm S}}^{\rm LE}, 
   f_{\mu, {\rm S}}^{\rm LE}, 
   f_{\tau, {\rm S}}^{\rm LE} \right)$                   &
   $\left( f_{e, {\rm S}}^{\rm HE}, 
   f_{\mu, {\rm S}}^{\rm HE}, 
   f_{\tau, {\rm S}}^{\rm HE} \right)$                   \\
   \hline
   PL                                                   &
   $8317.64_{-8001.41}^{+1454.73}$                     &
   $\left( 
   0.06_{-0.05}^{+0.57}, 
   0.94_{-0.57}^{+0.05}, 
   0
   \right)$                                             &
   $\left( 
   0.21_{-0.18}^{+0.47}, 
   0.79_{-0.47}^{+0.18}, 
   0
   \right)$                                            &
   200                                                 &
   $199.53_{-17.56}^{+24.35}$                          &
   $\left( 
   0.34_{-0.03}^{+0.03}, 
   0.66_{-0.03}^{+0.03}, 
   0 
   \right)$                                           &
   $\left( 
   0.00_{-0.00}^{+0.03}, 
   1.00_{-0.03}^{+0.00}, 
   0
   \right)$                                          \\
   &
   &
   &
   &
   $10^3$                                               &
   $933.25_{-101.49}^{+241.64}$                       &
   $\left( 
   0.34_{-0.02}^{+0.02}, 
   0.66_{-0.02}^{+0.02}, 
   0
   \right)$                                             &
   $\left(
   0.00_{-0.00}^{+0.04}, 
   1.00_{-0.04}^{+0.00}, 
   0
   \right)$                                             \\
   BPL                                                  &
   $97.72_{-20.10}^{+3138.21}$                        &
   $\left( 
   0.19_{-0.18}^{+0.43}, 
   0.81_{-0.43}^{+0.18}, 
   0 
   \right)$                                             &
   $\left( 
   0.02_{-0.00}^{+0.64}, 
   0.98_{-0.64}^{+0.00}, 
   0
   \right)$                                             &
   200                                                  &
   $199.53_{-4.54}^{+4.65}$                           &
   $\left( 
   0.34_{-0.04}^{+0.04}, 
   0.66_{-0.04}^{+0.04}, 
   0
   \right)$                                             &
   $\left( 
   0.00_{-0.00}^{+0.05}, 
   1.00_{-0.05}^{+0.00}, 
   0
   \right)$                                            \\
   &
   &
   &
   &
   $10^3$                                               &
   $1000.00_{-108.75}^{+96.48}$                       &
   $\left(
   0.34_{-0.02}^{+0.02}, 
   0.66_{-0.02}^{+0.02}, 
   0
   \right)$                                             &
   $\left( 
   0.00_{-0.00}^{+0.07}, 
   1.00_{-0.07}^{+0.00}, 
   0
   \right)$                                             \\
   Step                                                 &
   $79.43_{-19.18}^{+2009.86}$                        &
   $\left( 
   0.22_{-0.18}^{+0.43}, 
   0.78_{-0.43}^{+0.18}, 
   0
   \right)$                                             &
   $\left( 
   0.11_{-0.10}^{+0.53}, 
   0.89_{-0.53}^{+0.10}, 
   0
   \right)$                                             &
   200                                                  &
   $199.53_{-4.54}^{+4.65}$                           &
   $\left( 
   0.33_{-0.02}^{+0.02}, 
   0.67_{-0.02}^{+0.02}, 
   0
   \right)$                                             &
   $\left( 
   0.00_{-0.00}^{+0.02}, 
   1.00_{-0.02}^{+0.00}, 
   0
   \right)$                                             \\
   &
   &
   &
   &
   $10^3$                                               &
   $1000.00_{-22.76}^{+23.29}$                        &
   $\left( 
   0.33_{-0.02}^{+0.02}, 
   0.67_{-0.02}^{+0.02}, 
   0
   \right)$                                             &
   $\left( 
   0.00_{-0.00}^{+0.06}, 
   1.00_{-0.06}^{+0.00}, 
   0
   \right)$                                            \\
  \end{tabular}
 \end{ruledtabular}  
\end{table*}
\endgroup


\end{document}